\documentclass[compsoc,conference,a4paper,10pt,times]{IEEEtran}
\IEEEoverridecommandlockouts
\usepackage{cite}
\usepackage{booktabs}
\usepackage{amsmath,amssymb,amsfonts}
\usepackage{amsmath}
\usepackage{multirow}
\usepackage{algorithm}
\usepackage{algpseudocode}
\usepackage{graphicx}
\usepackage{textcomp}
\usepackage{bmpsize}
\usepackage{xcolor}
\usepackage{lipsum}
\usepackage[colorlinks=true,urlcolor=black]{hyperref}
\def\BibTeX{{\rm B\kern-.05em{\sc i\kern-.025em b}\kern-.08em
    T\kern-.1667em\lower.7ex\hbox{E}\kern-.125emX}}

\usepackage{listings}

\lstdefinestyle{customc}{
  belowcaptionskip=1\baselineskip,
  breaklines=true,
  frame=single,
  xleftmargin=0pt, 
  xrightmargin=0pt, 
  language={}, 
  showstringspaces=false,
  basicstyle=\footnotesize\ttfamily\color{black}, 
  moredelim=**[il][\color{pink}]{----}, 
  moredelim=**[il][\color{green}]{++++}, 
}

\begin{document}

\title{HALURust: Exploiting Hallucinations of Large Language Models to Detect Vulnerabilities in Rust}


\author{
\IEEEauthorblockN{Yu Luo}
\IEEEauthorblockA{
\textit{University of Central Missouri}\\
Warrensburg, MO, USA \\
yuluo@ucmo.edu}
\and
\IEEEauthorblockN{Han Zhou}
\IEEEauthorblockA{
\textit{University of Missouri-Kansas City}\\
Kansas City, MO, USA \\
hzb4f@umkc.edu}
\and
\IEEEauthorblockN{Mengtao Zhang}
\IEEEauthorblockA{
\textit{University of Missouri-Kansas City}\\
Kansas City, MO, USA \\
mengtaozhang@umkc.edu}
\and
\IEEEauthorblockN{Dylan De La Rosa}
\IEEEauthorblockA{
\textit{Texas A\&M University}\\
College Station, TX, USA \\
ddelarosa5@leomail.tamuc.edu}
\and
\IEEEauthorblockN{Hafsa Ahmed}
\IEEEauthorblockA{
\textit{University of Missouri-Kansas City}\\
Kansas City, MO, USA \\
habnz@umkc.edu}
\and
\IEEEauthorblockN{Weifeng Xu}
\IEEEauthorblockA{
\textit{The University of Baltimore}\\
Baltimore, MD, USA \\
wxu@ubalt.edu}
\and
\IEEEauthorblockN{Dianxiang Xu}
\IEEEauthorblockA{
\textit{University of Missouri-Kansas City}\\
Kansas City, MO, USA \\
dxu@umkc.edu}
}

\maketitle

\begin{abstract}
As an emerging programming language, Rust has rapidly gained popularity and recognition among developers due to its strong emphasis on safety. It employs a unique ownership system and safe concurrency practices to ensure robust safety. Despite these safeguards, security in Rust still presents challenges. Since 2018, 442 Rust-related vulnerabilities have been reported in real-world applications. The limited availability of data has resulted in existing vulnerability detection tools performing poorly in real-world scenarios, often failing to adapt to new and complex vulnerabilities. This paper introduces HALURust, a novel framework that leverages hallucinations of large language models (LLMs) to detect vulnerabilities in real-world Rust scenarios. HALURust leverages LLMs' strength in natural language generation by transforming code into detailed vulnerability analysis reports. The key innovation lies in prompting the LLM to always assume the presence of a vulnerability. If the code sample is vulnerable, the LLM provides an accurate analysis; if not, it generates a hallucinated report. By fine-tuning LLMs on these hallucinations, HALURust can effectively distinguish between vulnerable and non-vulnerable code samples. HALURust was evaluated on a dataset of 81 real-world vulnerabilities, covering 447 functions and 18,691 lines of code across 54 applications. It outperformed existing methods, achieving an F1 score of 77.3\%, with over 10\% improvement. The hallucinated report-based fine-tuning improved detection by 20\% compared to traditional code-based fine-tuning. Additionally, HALURust effectively adapted to unseen vulnerabilities and other programming languages, demonstrating strong generalization capabilities.
\end{abstract}


\section{Introduction}

Rust's design strategically focuses on safety and performance, targeting the elimination of common programming errors that lead to security vulnerabilities and system crashes. Its robust memory safety protocols and inherent defenses against data races demonstrate a strong commitment to security \cite{RustSecurity}, significantly boosting its popularity. Despite these safeguards, security in Rust is not without its challenges. According to the Common Vulnerabilities and Exposures (CVE) database \cite{CVE}, there have been 442 reported Rust-related vulnerabilities in real-world applications since 2018. Notably, in 2023, a severe vulnerability was discovered in Cargo, Rust's package manager \cite{Cargo}. This flaw involved Cargo's mishandling of `umask' settings during the extraction of crate archives on UNIX-like systems, which inadvertently permitted local users to modify the source code of these archives. Such modifications could lead to unauthorized code execution upon compilation and execution of the altered crates, highlighting a critical oversight in Rust's ecosystem. This gap between Rust's theoretical robustness and its practical vulnerabilities underscores the urgent need for effective vulnerability detection tools specifically tailored for Rust, to ensure continuous security enhancements and address emerging threats effectively.

Most existing research on Rust's safety primarily targets identifying safety-related bugs through various methods. For instance, Rudra \cite{bae2021rudra}, Safedrop \cite{cui2023safedrop}, and MirChecker \cite{li2021mirchecker} use static analysis, while Fidelius Charm \cite{almohri2018fidelius} and TRust \cite{bang2023trust} utilize the `unsafe' keyword methods. Crichton et al. \cite{crichton2023grounded, crichton2022modular} explore Rust's ownership mechanisms to enforce safe practices and manage memory and concurrency. Although these approaches effectively identify certain bugs and enforce safe programming practices, they are limited in detecting new, complex vulnerabilities that don't follow predefined patterns and require significant manual effort, which can be prone to human error.

In contrast, the adoption of deep learning (DL) for vulnerability detection in Rust is gaining traction due to its superior pattern recognition and adaptability to emerging vulnerabilities. Rustspot \cite{park2022unsafe} utilizes DL techniques, while Yuga \cite{nitin2023yuga} applies pattern-matching algorithms, and RustHorn \cite{matsushita2021rusthorn} employs constrained Horn clauses (CHCs) for detecting vulnerabilities in Rust. Additionally, some language-agnostic methods \cite{cao2022mvd, luo2022cag, wen2023vulnerability} employ Graph Neural Networks (GNNs) to enhance vulnerability detection. Moreover, the increasing use of large language models (LLMs) has further enhanced these capabilities. Models like CodeBERT \cite{feng2020codebert} and ChatGPT \cite{openai2024chatgpt} are being used in various studies \cite{cheshkov2023evaluation, le2024software,  pearce2023examining, zhang2023prompt} to automate the detection of vulnerabilities across a broad range of complexities and scopes.

Although these automated approaches offer significant advantages for detecting vulnerabilities in Rust, they still encounter several challenges as follows:

\begin{itemize}

\item \textbf{Limited Data.} Data quality and diversity are essential for effective pretraining and fine-tuning of AI models\cite{wang2023data}. As an emerging language with a strong emphasis on security, Rust presents challenges in gathering substantial and real vulnerability data. The limited availability of extensive and varied datasets hampers the models' capability to accurately detect vulnerabilities. Although Le et al. \cite{le2024software} contribute a dataset containing 90 vulnerable functions and 350 vulnerable lines, these are derived from just 31 CVE records. This means that despite the good volume of samples, they represent only 31 distinct vulnerabilities. Moreover, treating individual functions as separate vulnerable samples may not always be appropriate, as vulnerabilities often span multiple functions. This issue highlights that the availability of correct and comprehensive Rust vulnerability data is currently limited, impacting the development of effective vulnerability detection models.

\item \textbf{Limited Approaches.} Due to the limited availability of accurate and comprehensive Rust vulnerability data, there are relatively few methods specifically tailored for detecting vulnerabilities in this language. Most research in this area relies on empirical analyses to draw conclusions about bug or vulnerability detection in Rust. For instance, Qin et al. \cite{qin2020understanding}, Xu et al. \cite{xu2021memory}, and Zheng et al. \cite{zheng2023closer} have provided insights into the memory-safety issues associated with Rust, contributing to the ongoing development and refinement of the language. However, despite the prevalence of language-agnostic deep-learning-based approaches for vulnerability detection, there is a noticeable lack of research that discusses or adapts these methods specifically for Rust. 

\item \textbf{Narrow Focus.} Current approaches to vulnerability detection in Rust often exhibit a narrow focus, typically concentrating on a limited range of vulnerability types. This narrow scope results in poor scalability, making it challenging to adapt these methods to complex or emerging vulnerabilities. Furthermore, most of these approaches tend to offer only binary outcomes (vulnerable or non-vulnerable), without providing insights into the specific types of vulnerabilities present. This limitation reduces their effectiveness for detailed debugging and accurately pinpointing various vulnerabilities, thereby restricting their utility in comprehensive security analyses and remediation efforts.

\item \textbf{Poor Performance.} Existing vulnerability detection approaches in Rust often suffer from poor performance. Le et al. \cite{le2024software} conducted empirical experiments on real-world Rust vulnerability detection and demonstrated that two state-of-the-art methods, employing CodeBERT and ChatGPT, achieved F1 scores of less than 50\%. In contrast, many DL-based language-agnostic approaches report achieving over 90\% accuracy in detecting vulnerabilities. However, Chakraborty et al. \cite{chakraborty2021deep} evaluated multiple state-of-the-art deep learning-based techniques in real-world vulnerability prediction scenarios and reported a performance decline exceeding 50\%. Further, Ullah et al. \cite{ullah2024llms} conclude that state-of-the-art LLMs are not yet ready to be used for vulnerability detection. These findings underscore the challenges that current methods face in consistently and accurately identifying vulnerabilities in Rust code under practical conditions.

\end{itemize}

This paper aims to address the challenges encountered in detecting vulnerabilities within Rust code. In particular, our contributions are as follows:


\begin{itemize}

\item We have developed a Rust vulnerability dataset based on 81 CVE records from real-world applications, covering 44 different Common Weakness Enumerations (CWEs) \cite{CWE}. For each CVE record, we extract all vulnerability-related functions as a single vulnerable sample, and the corresponding fix code as a non-vulnerable sample. In total, this dataset includes 447 functions and 18,691 lines of code, spanning 54 different programs.

\item We propose HALURust, a novel framework that leverages hallucinations of LLMs to detect vulnerabilities in real-world Rust scenarios. In the context of language models, ``hallucination'' refers to the generation of incorrect or misleading information that the model presents as factual. While many studies \cite{agrawal2023can, ji2023towards, liang2024learning} focus on mitigating hallucinations, HALURust uniquely leverages this ``bad'' characteristic to its advantage. Specifically, we prompt the LLM to assume that every code sample contains a vulnerability. For vulnerable code, this assumption is a truth, leading the LLM to generate an accurate analysis report. For non-vulnerable code, the assumption becomes a lie, resulting in a hallucinated report. By fine-tuning LLMs on both real and hallucinated reports, HALURust can detect hallucinations in the report, thereby achieving differentiation between vulnerable and non-vulnerable code samples.

\item HALURust employs multiple LLMs, including GPT-3.5 Turbo \cite{OpenAIgpt3.5}, GPT-4o \cite{OpenAIgpt4o}, Llama3 \cite{llama3modelcard}, Mistral \cite{Mistral}, and Codellama \cite{roziere2023code} to produce vulnerability analysis reports, which expands the range of text-based report data available for model training. To avoid overfitting caused by hallucinations that are easily learned by the same model during training, these reports are then used to fine-tune a different LLM, Gemma-7b \cite{team2024gemma}, aiming to improve the model's ability to accurately identify vulnerabilities. An ablation study has demonstrated the superior performance of fine-tuning with report formats over direct code formats in vulnerability detection.

\item We evaluated HALURust on a diverse range of real-world code samples. The empirical findings indicate that HALURust surpasses existing state-of-the-art methods, achieving the highest accuracy of 77.4\% and F1 scores of 77.3\%, which represents an improvement of over 10\%. Furthermore, the innovative hallucinated report-based fine-tuning method increased detection performance by 20\% compared to traditional code-based fine-tuning. Additionally, when applied to unseen types of vulnerabilities, HALURust also demonstrated commendable performance, achieving an accuracy of approximately 65\%.

\end{itemize}

The remainder of this paper is organized as follows. Section 2 reviews related work; Section 3 describes vulnerabilities in Rust; Section 4 introduces the architecture of HALURust for vulnerability detection; Section 5 presents the experiment results; Section 6 discusses the threats to validity; Section 7 concludes this paper.

\section{Related Work}

The majority of existing research on Rust's safety primarily focuses on identifying safety-related bugs. Rudra \cite{bae2021rudra} analyzes all packages hosted in the Rust package registry to detect memory safety bugs on a large scale across the Rust ecosystem. Safedrop \cite{cui2023safedrop} targets invalid memory deallocation, leveraging a modified Tarjan algorithm for scalable path-sensitive analysis combined with a cache-based strategy to detect such bugs. MirChecker \cite{li2021mirchecker} employs static analysis on Rust’s Mid-level Intermediate Representation (MIR) to detect potential runtime crashes and memory-safety errors. Fidelius Charm \cite{almohri2018fidelius} and TRust \cite{bang2023trust} use the `unsafe' keyword to isolate potentially unsafe Rust code, thereby reducing the volume of code that could lead to memory safety issues. Additionally, Crichton et al. \cite{crichton2023grounded, crichton2022modular} explore Rust's ownership mechanisms to safely and efficiently manage memory and concurrency, aiming to mitigate the risks of commonly encountered bugs and vulnerabilities in systems programming.


Various advanced methods are employed to ensure Rust's safety. Rustspot \cite{park2022unsafe} employs ML techniques to detect memory-safety bugs in Rust's binary code. Yuga \cite{nitin2023yuga} enhances bug identification through rapid pattern-matching algorithms and field-sensitive alias analysis. RustHorn \cite{matsushita2021rusthorn} advances the verification of Rust programs for functional correctness by transforming them into constrained horn clauses (CHCs). RustBelt \cite{jung2017rustbelt} provides the first formal and machine-checked safety proof for a language representing a realistic subset of Rust, while RustHornBelt \cite{matsushita2022rusthornbelt} extends this work by giving first-order logic (FOL) specifications to safe APIs implemented with unsafe code. Additional studies \cite{astrauskas2020programmers, evans2020rust, qin2020understanding} analyze how unsafe code is used in practice and the purpose of unsafe code to enhance Rust's safety.

The rising popularity of large language models (LLMs) has also expanded the possibilities for vulnerability detection, automating the identification of vulnerabilities across diverse scenarios. Le et al. \cite{le2024software} utilize models such as CodeBERT \cite{feng2020codebert} and ChatGPT \cite{openai2024chatgpt} to identify vulnerabilities at both the line and function levels in Rust code. Cheshkov et al. \cite{cheshkov2023evaluation} and Zhang et al. \cite{zhang2023prompt} have effectively used ChatGPT, with the latter enhancing detection performance through strategic prompt engineering. Moreover, Pearce et al. \cite{pearce2023examining} have assessed the effectiveness of various LLMs in addressing software vulnerabilities. Ullah et al. \cite{ullah2024llms} applied various prompt engineering techniques, including task-oriented and role-oriented prompts, with both zero-shot and few-shot settings, to evaluate LLMs for vulnerability detection.

Another aspect of related work is the exploration of controlling hallucinations in LLMs. Yao et al. \cite{yao2023llm} argued that hallucinations can sometimes be beneficial rather than problematic, depending on the context in which they are used. Sui et al. \cite{sui2024confabulation} highlighted that hallucinations are useful for producing narrative-rich and coherent outputs, especially in scenarios that require storytelling or creative improvisation. ProMaC \cite{hu2024leveraging} also leverages hallucinations by expanding plausible contexts and iteratively refining prompts for segmentation tasks, showing that hallucinations can play a constructive role in enhancing task-specific performance. On the other hand, several studies \cite{mundler2023self, zhang2023language, jones2023teaching} focus on understanding the impact of hallucinations and developing mitigation techniques.

\section{Vulnerability in Rust}

As a modern language that balances high performance with robust security features, Rust ensures safety through its unique ownership system and stringent compile-time checks that help prevent common programming errors. The ownership system enforces strict rules on how data is managed in memory. Every piece of data has a clear owner, and its lifetime is explicitly tied to the scope of that owner, ending when the owner goes out of scope. Additionally, Rust’s borrowing mechanism allows other parts of the code to temporarily access data without assuming full ownership. This structure not only prevents data races and dangling references but also maintains the integrity of data handling throughout the program’s execution.

Despite these comprehensive safeguards, vulnerabilities in Rust can still occur, particularly when developers need to bypass the compiler's strict safety guarantees. Rust provides the `unsafe' keyword to allow developers to perform low-level operations or optimize performance by manually managing memory and other system resources. These `unsafe' blocks open the door to potential risks such as undefined behaviors and memory safety bugs. Empirical studies \cite{cui2023safedrop, zheng2023closer} have explored potential vulnerabilities by examining unsafe API usage and the characteristics of disclosed vulnerabilities within the Rust ecosystem.

\begin{table}[h]
\caption{Categories of CWE}
\begin{center}
\begin{tabular}{|l|l|}
\hline
\textbf{Category}&\textbf{CWEs}\\
\hline
& CWE-119, CWE-125, CWE-131\\
Memory Safety & CWE-190, CWE-191,  CWE-415\\
& CWE-416, CWE-475, CWE-665\\
& CWE-787, CWE-824, CWE-908\\
\hline
 & CWE-020, CWE-022, CWE-059\\
& CWE-079, CWE-088, CWE-113\\
Input Validation & CWE-134, CWE-200, CWE-203\\
& CWE-287, CWE-288, CWE-346\\
& CWE-427, CWE-444\\
\hline
& CWE-276, CWE-362, CWE-400\\
Concurrency Issue & CWE-617, CWE-662, CWE-674\\
& CWE-668, CWE-703, CWE-770\\
\hline
& CWE-248, CWE-252, CWE-347\\
Security Handling & CWE-670, CWE-682, CWE-701\\
& CWE-754, CWE-755, CWE-758\\

\hline
\end{tabular}
\label{table:CWEGroup}
\end{center}
\end{table}

Based on the nature of the vulnerabilities and the aspects of system security they impact, we have categorized all CWEs present in our dataset into the following four groups, as detailed in table \ref{table:CWEGroup}.

\begin{figure*}[h]
  \centering
  \includegraphics[width = 0.96\linewidth, height = 7cm]{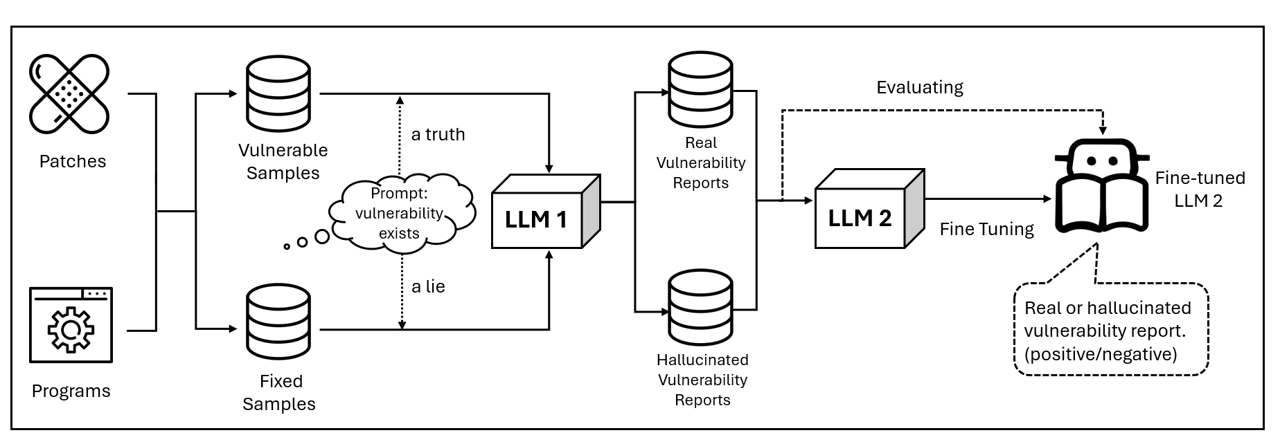}
  \caption{The Architecture of HALURust for Predicting Vulnerabilities in Rust}
  \label{fig: Architecture}
\end{figure*}

\begin{itemize}

\item \textbf{Memory-safety Vulnerability.} Memory safety vulnerabilities typically arise from improper handling of memory operations, including allocation, deallocation, and access. These mishandlings can lead to critical issues such as buffer overflows, memory leaks, or use-after-free errors. For example, CWE-119 involves the improper restriction of operations within the bounds of a memory buffer. This occurs when a program writes outside the boundaries of an allocated buffer, potentially leading to code execution or denial of service. Similarly, CWE-416 refers to use-after-free errors, which occur when a program continues to use a pointer after it has been freed, leading to potential crashes or code execution.

\item \textbf{Input Validation Vulnerability.} Input validation vulnerabilities occur when a program fails to properly validate or sanitize input from users or other external sources, This oversight can lead to various security issues, including injection attacks, buffer overflows, and information disclosure. For example, CWE-113 involves improper neutralization of CRLF sequences in HTTP headers. This occurs when input is improperly sanitized, allowing attackers to manipulate HTTP headers.

\item \textbf{Concurrency Vulnerability.} Concurrency issues arise when multiple threads or processes access shared resources without proper synchronization, leading to race conditions, deadlocks, and other unpredictable behaviors. For example, CWE-362 refers to concurrent execution using shared resources with improper synchronization(`Race Condition'). This involves multiple threads or processes accessing shared resources without proper synchronization, leading to race conditions.

\item \textbf{Security Handling Vulnerabily.} Security handling vulnerabilities involve improper implementation of security features, leading to potential bypass of security controls, exposure of sensitive data, and other security failures. For example, CWE-701 involves improper initialization of variables or states, leading to undefined behavior.


\end{itemize}

\section{Methodology}

Figure \ref{fig: Architecture} shows the architecture of HALURust for predicting vulnerabilities in Rust code by leveraging historical data from the Common Vulnerabilities and Exposures (CVE) dataset across three steps:

\begin{itemize}

\item \textbf{Step 1.} For each CVE record, we systematically extract the patch associated with the reported vulnerability. This patch provides critical information including the affected program version, the specific files impacted, and segments of both vulnerable and fixed code. Utilizing this data, we identify the relevant program version and precisely locate the files containing vulnerabilities. We further extract entire functions that encompass the vulnerable code segments to construct a vulnerable sample. Correspondingly, functions containing the fixed code are extracted as a non-vulnerable sample.

\item \textbf{Step 2.} Continuing from the initial data preparation, we prompt LLMs to generate vulnerability analysis reports. In the prompt, we make a critical assumption: each code sample contains a vulnerability, regardless of whether it was originally identified as vulnerable or non-vulnerable. This approach allows LLMs to produce detailed reports for each sample, attempting to identify and analyze any vulnerabilities present. For samples that truly contain vulnerabilities, the prompt aligns with reality, and the LLM generates accurate vulnerability analyses. However, for non-vulnerable samples, the prompt becomes a lie, leading the LLM to produce reports on non-existent vulnerabilities, which we refer to as hallucinated vulnerability reports.

\item \textbf{Step 3.} These generated reports, comprising both real and hallucinated vulnerability reports, serve as intermediate textual representations for further analysis. We divide these reports into two datasets: one for fine-tuning and the other for evaluation. To avoid the bias where an LLM can easily identify hallucinated features it generated itself, we use these samples to fine-tune a different LLM. This new model differentiates between reports that accurately describe real vulnerabilities and those that erroneously report vulnerabilities where none exist. For each new sample, the classification outcome of its report determines its status: if the report is identified as a real vulnerability report, then the sample is considered vulnerable. Conversely, if the report is classified as a hallucinated vulnerability report, indicating the analysis of non-existent vulnerabilities, the sample is deemed non-vulnerable.

\end{itemize}

The details of each step are
further elaborated as follows.

\subsection{Data Preparation}

Data preparation involves an extensive extraction and categorization process based on the CVE dataset. Each CVE record, which contains data about known vulnerabilities, includes information about the associated patch detailing the software version, the specific files affected, and the code segments that are vulnerable and have been fixed. List \ref{lst:CVE} displays a CVE record \cite{CVESample1} created in 2018 with the ID CVE-2018-1000657. It identifies a buffer overflow vulnerability in the Rust standard library's `VecDeque::reserve()' function, which affects Rust versions starting from the stable release 1.3.0. This vulnerability could potentially allow arbitrary code execution, although no proof-of-concept exploit has been published. The vulnerability was rectified in versions starting from stable release 1.22.0 onwards.

\begin{lstlisting}[style=customc, caption=CVE-2018-1000657, captionpos=b, label=lst:CVE]
CVE-2018-1000657
--
Description:
Rust Programming Language Rust standard library version stable release 1.3.0 and later contains a Buffer Overflow vulnerability in std::collections::vec_deque::VecDeque::reserve() function that can result in Arbitrary code execution, but no proof-of-concept exploit is currently published. This vulnerability appears to have been fixed after the stable release 1.22.0 and later.
--
src/liballoc/vec_deque.rs
--
@@ -558,7 +558,7 @@ impl<T> VecDeque<T> {
           .and_then(|needed_cap| needed_cap.checked_next_power_of_two())
           .expect("capacity overflow");
-----     if new_cap > self.capacity() {
+++++     if new_cap > old_cap {
           self.buf.reserve_exact(used_cap, new_cap - used_cap);
           unsafe {
               self.handle_cap_increase(old_cap);
\end{lstlisting}

Specifically, the vulnerability is located in the file `src/liballoc/vec\_deque.rs' within the Rust standard library. This file contains the implementation of the `VecDeque' data structure. The vulnerability arises in the `VecDeque::reserve()' function, where the original code contains a conditional statement if `new\_cap $>$ self.capacity()' marked by a pink line starting with `-'. This statement checks whether the newly calculated capacity `new\_cap' for a `VecDeque' exceeds its current capacity. However, if there is a discrepancy between the logical capacity (what the code believes is allocated) and the physical capacity, such handling of the capacity check is flawed and could potentially lead to a buffer overflow. To rectify this issue, the patch modifies the conditional check to `if new\_cap $>$ old\_cap', highlighted by a green line starting with `+'. This change ensures that `new\_cap' is compared against the previous capacity `old\_cap', rather than the VecDeque's current capacity `self.capacity()'. This change addresses the risk of a buffer overflow by properly managing the conditions under which the VecDeque's capacity is increased.

\begin{lstlisting}[style=customc, caption=Generated Samples on CVE-2018-1000657, captionpos=b, label=lst:CVESample]
Vulnerable Sample:
pub fn reserve(&mut self, additional: usize) {
    let old_cap = self.cap();
    let used_cap = self.len() + 1;
    let new_cap = used_cap.checked_add(additional)
        .and_then(|needed_cap| needed_cap.checked_next_power_of_two())
        .expect("capacity overflow");
----    if new_cap > self.capacity() {
        self.buf.reserve_exact(used_cap, new_cap - used_cap);
        unsafe {
            self.handle_cap_increase(old_cap);
        }
    }
}
Non-vulnerable Sample:
pub fn reserve(&mut self, additional: usize) {
    let old_cap = self.cap();
    let used_cap = self.len() + 1;
    let new_cap = used_cap.checked_add(additional)
        .and_then(|needed_cap| needed_cap.checked_next_power_of_two())
        .expect("capacity overflow");
++++    if new_cap > old_cap {
        self.buf.reserve_exact(used_cap, new_cap - used_cap);
        unsafe {
            self.handle_cap_increase(old_cap);
        }
    }
}
\end{lstlisting}

However, while the patch outlines the changes from vulnerable to fixed code, it lacks detailed declarations such as that of `old\_cap', as well as other structural information critical for comprehensive static analysis and fine-tuning LLMs. These models require a deep understanding of the function structure and interrelationships within the software. To address this, we systematically track specific versions of the program and combine all related functions into a single, informative sample. List \ref{lst:CVESample} illustrates how we generate two distinct samples based on the CVE record: a vulnerable sample containing only the problematic code and a non-vulnerable sample containing only the corrected code. The vulnerable function `reserve' triggers a buffer overflow by comparing `new\_cap' with the current capacity `self.capacity()', whereas in the fixed version, the comparison is made against `old\_cap', ensuring safer capacity management.

This approach effectively generates one vulnerable and one non-vulnerable sample from each CVE record, establishing a comprehensive dataset for further analysis and model training. Notably, in the example provided, the patch resolves the vulnerability by modifying a single function. However, more intricate patches might address vulnerabilities through changes spanning multiple interconnected functions. As a result, each sample in our dataset could consist of one or more functions, along with any relevant global variables. This flexibility indicates that the inputs for HALURust are not confined to a single function but could include multiple functions, a class, or even an entire code segment. This versatility enhances HALURust's capability to handle complex and varied software security scenarios.

\subsection{Reports Generation}

In real-world software development, vulnerable code can significantly vary in length, span multiple functions or classes, and relate to different project scenarios. These variations present significant challenges for the detection of security vulnerabilities through static analysis, particularly in the context of emerging programming languages where available data is limited. Furthermore, Chakraborty et al. \cite{chakraborty2021deep} assess several state-of-the-art deep-learning-based analysis methods in actual vulnerability prediction scenarios and report a performance decrease of over 50\%. Ullah et al. \cite{ullah2024llms} concluded that directly using LLMs to detect vulnerabilities from code input does not yield significant effectiveness.

\begin{lstlisting}[style=customc, caption= The  Real Vulnerability Report, captionpos=b, label=lst:CVEReport]
----Capacity Check:
The function is comparing new_cap against self.capacity(). However, self.capacity() returns the current buffer capacity, which might not have been updated yet. This can result in a mismatch between the actual buffer size and what is checked before allocating more space. If the buffer is already too small but self.capacity() inaccurately reports a larger size (due to a previous reallocation issue, for example), then no new allocation will occur, and further writes could overflow the buffer.

----Unchecked Assumption:
The unsafe block assumes that old_cap is valid for handling the buffer reallocation, but if the comparison (new_cap > self.capacity()) fails due to an outdated or inaccurate capacity, then you could end up using memory space that hasn't been properly allocated, causing a buffer overflow.
\end{lstlisting}

To address these challenges, we harness the robust capabilities of LLMs to transform complex code data into text-based report formats. List \ref{lst:CVEReport} shows a real vulnerability report generated by GPT-4o, analyzing the vulnerable sample from list \ref{lst:CVESample}. This report not only accurately identifies the comparison of `new\_cap' against `self.capacity()' as a potential cause of buffer overflow but also explains the flawed logic in the capacity calculation that leads to the vulnerability. In contrast, list \ref{lst:CVEReport} displays a hallucinated vulnerability report analyzing the non-vulnerable sample from list \ref{lst:CVESample}. It suggests that `self
.handle\_cap\_increase(old\_cap)' is called without any visible checks, while in reality, an if condition checks `new\_cap' against `old\_cap' to ensure proper memory allocation.

Generally, the report format can specify and localize complex code details within a constrained scope, offering clear conclusions and pinpointing specific code sections with supporting evidence for comprehensive analysis. The subsequent classification task can be accomplished by assessing the reasonableness and alignment of the statements and evidence presented in the reports. Our approach operates under the assumption that each sample in our dataset exhibits a vulnerability. The process is structured into three steps: (1) design custom prompts specific to each scenario, (2) utilize multiple LLMs to analyze each sample and generate comprehensive reports, and (3) assign a binary label (positive or negative) to each sample.

\begin{lstlisting}[style=customc, caption= The  Hallucinated Vulnerability Report, captionpos=b, label=lst:HaluCVEReport]
----Unchecked unsafe block: 
The unsafe block in the function is calling self.handle_cap_increase(old_cap) without any visible checks. Since Rust's unsafe code bypasses some of the language's safety guarantees (like memory safety), it is crucial to ensure that handle_cap_increase is implemented safely. If this function contains unsafe memory handling (e.g., incorrect pointer manipulation or improper use of allocations), it could lead to vulnerabilities such as memory corruption, undefined behavior, or use-after-free issues.

\end{lstlisting}

\textbf{Prompts.} To facilitate the interaction with LLMs for producing informative and accurate reports, we apply three different techniques, task-oriented (TO) \cite{brown2020language}, role-oriented (RO) \cite{zhou2022large}, and CO-STAR \cite{COSTAR}, ranging from simple to complex, and demonstrate in subsequent experiments how different prompts generate varying vulnerability reports that impact model performance.

A TO prompt provides clear instructions for a specific task that the LLM must perform. It focuses the model’s output towards achieving a well-defined goal. List \ref{lst:TOPrompt} shows the TO prompt used in HALURust, which instructs the model to perform a comprehensive analysis of Rust code suspected of containing vulnerabilities and to generate a concise technical report with a word limit of 500. Notably, for vulnerable samples, the context is the description derived from each CVE record, which guides the report to accurately address real vulnerabilities and ensures correct labeling for subsequent training. For non-vulnerable samples, where a vulnerability is assumed, the context highlights the absence of detailed project information or historical issues, emphasizing the need for a thorough and focused review of the Rust code to uncover potential vulnerabilities. In testing samples, the latter context is used to maintain fairness across evaluations.

\begin{lstlisting}[style=customc, caption= The Task-Oriented Prompt, captionpos=b, label=lst:TOPrompt]
The Rust code is suspected to have a vulnerability, but no specific project details or past issues are known.  
OR CVE description.
Perform a comprehensive analysis of the Rust code, aiming to identify sections that are most likely to contain critical vulnerabilities. The goal is to thoroughly scan the entire codebase and provide a detailed technical report, with a word limit of 500 words, highlighting potential vulnerabilities.

\end{lstlisting}

An RO prompt assigns a specific role to the LLM to guide its output, helping the model adopt a particular perspective and improving contextual relevance. List \ref{lst:ROPrompt} shows the RO prompt used in HALURust. Based on the TO prompt, the model is instructed to act as a "Rust security expert" and perform a comprehensive analysis of the code to identify potential vulnerabilities.

\begin{lstlisting}[style=customc, caption= The Role-Oriented Prompt, captionpos=b, label=lst:ROPrompt]
The Rust code is suspected to have a vulnerability, but no specific project details or past issues are known.  
OR CVE description.
I want you to act as a Rust security expert and perform a comprehensive analysis of the Rust code, aiming to identify sections that are most likely to contain critical vulnerabilities. The goal is to thoroughly scan the entire codebase and provide a detailed technical report, with a word limit of 500 words, highlighting potential vulnerabilities.

\end{lstlisting}

CO-STAR, a widely used prompt framework, ensures that all key aspects influencing an LLM's response are meticulously considered, yielding more tailored and impactful outcomes. The constructed prompt, as illustrated in list \ref{lst:Prompt}, encompasses six critical aspects.

\begin{lstlisting}[style=customc, caption= The CO-STAR Prompt, captionpos=b, label=lst:Prompt]
#Context#
The Rust code is suspected to have a vulnerability, but no specific project details or past issues are known. This absence of context necessitates a comprehensive and focused review to identify the most significant security concern, covering a wide range of common vulnerabilities in Rust.
OR CVE description.
#Objective#
The objective is to perform a detailed scan of the entire code to pinpoint the section most likely to contain a critical vulnerability.
#Style#
Request a detailed analysis in a technical report format. The report should focus exclusively on the most significant vulnerable code snippet, annotated with comments that explain why this section is particularly problematic.
#Tone#
Adopt a professional and investigative tone, reflecting the systematic approach required to uncover and document the most critical vulnerability.
#Audience#
The intended audience is people who have a good understanding of Rust and are familiar with common vulnerabilities. The report should provide insights that are technically thorough yet accessible to these professionals.
#Response#
The response should be a structured report that focuses solely on the most critical vulnerability identified, with a word limit of 500 words.  It should only provide a clear description of the vulnerability and its location in the code. The report should not include any recommendations for mitigation, impact, conclusion, or suggestion.
\end{lstlisting}

\begin{itemize}
\item \textbf{Context} - Same as the context in TO and RO prompts, ensuring correct labeling.

\item \textbf{Objective} - The objective is to perform a detailed scan of the entire code to identify the section most likely to contain a critical vulnerability. This includes generating hallucination vulnerabilities for non-vulnerable samples for analysis purposes and providing accurate analyses for vulnerable samples.

\item \textbf{Style} - The report is required to adopt a technical format, focusing exclusively on significant vulnerabilities and including annotations that explain why these code snippets are deemed vulnerable. This is designed to facilitate the analysis of the relationships among the vulnerability, the vulnerable code, and the explanations provided, enhancing the model's ability to predict reasonableness.

\item \textbf{Tone} - A professional and investigative tone is set up, appropriate for systematic examination to uncover and document vulnerabilities within the code.

\item \textbf{Audience} - The intended audience consists of professionals familiar with Rust and its common vulnerabilities, requiring a presentation that is both technically thorough and accessible. This approach not only facilitates the AI models' ability to distinguish between actual vulnerabilities and hallucinations in the reports but also ensures that the content is easily understandable and useful for human experts.

\item \textbf{Response} - The response should be a structured report focused on identifying and describing the most critical vulnerability, with a word limit of 500 words. It will exclude recommendations for mitigation, implications of the vulnerability, or any conclusions beyond the identification and description of the vulnerability itself. Including additional content could detract from the clarity and focus required for subsequent analysis by AI models.
\end{itemize}

\textbf{LLMs for Report Generation.} We utilize five distinct LLMs to generate vulnerability analysis reports. The models selected for this task are GPT-3.5 Turbo \cite{OpenAIgpt3.5}, GPT-4o \cite{OpenAIgpt4o}, Llama3 \cite{llama3modelcard}, Mistral \cite{Mistral}, and Codellama \cite{roziere2023code} each known for their robust performance across multiple benchmark datasets. GPT-3.5 Turbo and GPT-4o are known for their rapid response capabilities and high accuracy. It excels in generating detailed and contextually relevant content quickly. Llama3 stands out for its ability to adapt to various linguistic contexts, making it especially effective at producing clear and reliable reports from complex technical data. Mistral is adept at crafting coherent, long-form content, maintaining precision and clarity in extended narratives essential for detailed vulnerability analysis. Codellama is designed to assist with coding and programming tasks by providing solutions, explanations, and error analysis to enhance productivity and efficiency in software development.

\textbf{Labels.} Labeling samples is essential for the supervised fine-tuning of an LLM, as these labels provide critical information that the model leverages to learn the specified tasks. Reports derived from vulnerable samples that detail real vulnerabilities are labeled as `positive', while reports based on fixed samples, which include hallucinated vulnerabilities, are labeled as `negative'.

\subsection{Fine-tune LLMs}

Before fine-tuning LLMs, the dataset is divided into two subsets: one for fine-tuning the LLM and the other for evaluation purposes. To enhance the diversity and improve the generalizability of the dataset employed during fine-tuning, we utilize cosine similarity \cite{CosineSimilarity} as a measure. This metric is defined in Equation 1, where \( \mathbf{U} \) and \( \mathbf{W} \) are encoded vectors of two samples, \( \cdot \) represents the dot product, and \( \| \| \) denotes the magnitude of a vector. These encoded vectors are derived using the pre-trained model Codellama \cite{roziere2023code}.

\begin{equation}
\text{sim}(\mathbf{U}, \mathbf{W}) = \frac{\mathbf{U} \cdot \mathbf{W}}{\|\mathbf{U}\| \|\mathbf{W}\|}
\end{equation}

\begin{algorithm}
\caption{Diverse Sample Selection}
\label{alg:Selection}
\begin{algorithmic}[1]
\Require Set of samples $S$, Cosine\_similarity function $sim()$, Encoder $E()$, Select proportion $p$
\Ensure Diverse subset $S'$ 
\State $S' \leftarrow \emptyset$
\State $Dict \leftarrow (\{s : E(s)\} \mid s \in S)$
\State Randomly select $x$, where $x \in S$
\State $S' \leftarrow S' \cup \{x\}$
\State $S \leftarrow S \setminus \{x\}$
\While{$|S'| < p \times |S|$}
    \State min\_sim $\gets 1$
    \State candidate $\gets null$
    \For{$s \in S$}
        \State similarity $\gets sim(Dict[S'[-1]], Dict[s])$
        \If{ similarity $< min\_sim$}
            \State $min\_sim \leftarrow similarity$
            \State candidate $\leftarrow s$
        \EndIf
    \EndFor
    \State $S' \leftarrow S' \cup$ candidate
    \State $S \leftarrow S \setminus \{\text{candidate}\}$
\EndWhile
\State \Return $S'$
\end{algorithmic}
\end{algorithm}

Algorithm \ref{alg:Selection} outlines the process for selecting a diverse subset of samples from a dataset $S$, where $sim()$ calculates the cosine similarity between two vectors, and $E()$ encodes a sample into a vector using Codellama. The selection proportion $p$ dictates the fraction of the original set $S$ to be included in the subset $S'$. Initially, $S'$ is empty. All samples in $S$ are encoded and stored in a dictionary `Dict' to avoid redundant encoding. A random sample $x$ is chosen from $S$, added to $S'$, and removed from $S$. The selection continues in a while loop, which executes until the size of $S'$ reaches $p$ times the initial size of $S$. Within each iteration of the loop, the algorithm initializes `min\_sim' to 1 and `candidate' to null. For each sample $s$ in $S$, it computes the cosine similarity between the vector of the most recently added sample to $S'$ and the vector of $s$. If this similarity is less than `min\_sim', `min\_sim' is updated to this new similarity value, and `candidate' is updated to $s$. After each iteration, the identified `candidate' is added to $S'$ and removed from $S$, ensuring no sample is considered more than once. The process repeats until the desired proportion of samples is selected. 

The selected samples with binary labels serve as training data for fine-tuning. To avoid fine-tuning the same LLM used for report generation, which could lead to overfitting due to certain features being easier to learn, we use a different LLM, Gemma-7B \cite{team2024gemma}. This process employs the Low-Rank Adaptation (LoRA) method \cite{hu2021lora}, which specifically targets the \( q\_\text{proj} \) and \( v\_\text{proj} \) parameters to adjust the model architecture minimally while maximizing learning efficiency from the new data.

LoRA essentially modifies the weight matrix of a neural network by adding a low-rank update. Given an original weight matrix \( W \in \mathbb{R}^{d \times k} \), LoRA decomposes the update into two smaller matrices, \( A \in \mathbb{R}^{d \times r} \) and \( B \in \mathbb{R}^{r \times k} \), where \( r \ll d, k \). The effective weight after adaptation is given by:
\[
W' = W + AB
\]
This approach reduces the number of trainable parameters significantly, as it focuses on learning the low-rank components, making the process efficient for fine-tuning large models without altering the entire architecture.

The fine-tuning utilizes a cosine learning rate scheduler that begins at a learning rate of $1 \times 10^{-6}$ and incorporates a weight decay of $1 \times 10^{-4}$ to prevent overfitting. Training is managed with a small batch size of 2, an approach complemented by gradient accumulation over 2 steps. This strategy is designed to balance the computational demands with the limited memory capacity typically available, enabling the indirect handling of larger batches. Additionally, the training incorporates mixed precision techniques (FP16) \cite{micikevicius2017mixed} to enhance computational efficiency.

The remaining samples are used to evaluate the fine-tuned LLM. For classification, the model predicts each sample as either positive or negative, with correct predictions matching the ground truth labels.

\section{Experiment}

We utilized three distinct prompts to engage LLMs in generating reports through Ollama \cite{ollama} and Python 3.12.3 \cite{Python}. Additionally, we fine-tuned the Gemma-7B model using Pytorch v2.2.0 \cite{Pytorch}, Transformers v4.39.3 \cite{wolf2020transformers}, and LLaMA-Factory \cite{zheng2024llamafactory}. Our experiments were conducted on a multi-core server equipped with four Tesla V100S-PCIE GPUs. To evaluate the performance of HALURust, we employed common metrics such as accuracy, precision, recall, and the F1 score. These metrics are derived from the counts of true positives ($TP$), true negatives ($TN$), false positives ($FP$), and false negatives ($FN$). Accuracy is calculated as the proportion of correctly predicted samples (both positive and negative) out of the total samples, expressed as $(TP+TN)/(TP+TN+FP+FN)$. Precision denotes the ratio of correctly predicted positive samples to all predicted positive samples, i.e., $TP/(TP+FP)$. Recall is the fraction of actual positive samples that are correctly identified, i.e., $TP/(TP+FN)$. The F1 score, which is the harmonic mean of precision and recall, serves as a balanced measure of model performance, calculated as $2*(precision*recall)/(precision+recall)$. Collectively, these metrics provide a comprehensive evaluation of the effectiveness of a vulnerability detection model by capturing different aspects of its performance.

We employed 5 rounds of sample selection on the dataset of each experiment, allocating 80\% of the samples for fine-tuning LLMs and the remaining 20\% for evaluation. The initial epoch for each experiment was set at 350. If the accuracy and loss values were satisfactory after 350 epochs, we stopped the training. Otherwise, we continued with further training. Our reported evaluation metrics are based on the geometric mean with the maximum upward and downward deviations across the 5 selections.

\subsection{Dataset}

\begin{table}[htbp]
\caption{The Rust Dataset}
\begin{center}
\label{tab:data}
\begin{tabular}{|c|c|c|c|c|}
\hline
\#CVE Records & \#Functions &  LOC &\#CWEs & \#Programs \\
\hline
81 & 447 & 18,691 & 44 & 54\\ 
\hline
\end{tabular}
\end{center}
\end{table}

All samples in our dataset are derived from real-world applications, as detailed in table \ref{tab:data}. The dataset includes data extracted from 81 CVE records, encompassing 447 functions and 18,691 lines of code across 54 different programs. Collectively, these samples encompass 44 different types of CWEs, with detailed counts presented in table \ref{table:CWEs}. Notably, memory-related issues such as Improper Restriction of Operations within the Bounds of a Memory Buffer (CWE-119), Uncontrolled Resource Consumption (CWE-400), and Use After Free (CWE-416) are among the most frequently occurring vulnerabilities.  Details about each specific CVE record, including the ID, CWE type, program, and reference, are provided in table \ref{tab:CVEInfo} and \ref{tab:CVEInfo2} in Appendix A. The dataset and source code for reproduction are available at \cite{EURO_SP_Dataset}.

\begin{table}[htbp]
\caption{Numbers of Different CWEs}
\begin{center}
\begin{tabular}{|c|c|c|c|c|c|}
\hline
\textbf{CWEs}& \textbf{Num.} & \textbf{CWEs} & \textbf{Num.} &\textbf{CWEs}& \textbf{Num.}   \\
\hline
CWE-020 & 4 & CWE-252 & 2 & CWE-665 & 1 \\
\hline
CWE-022 & 2 & CWE-276 & 1 & CWE-668 & 1 \\
\hline
CWE-059 & 2 & CWE-287 & 1 & CWE-670 & 2 \\
\hline
CWE-079 & 1 & CWE-288 & 1 & CWE-674 & 1 \\
\hline
CWE-088 & 1 & CWE-346 & 1 & CWE-682 & 2 \\
\hline
CWE-113 & 1 & CWE-347 & 1 & CWE-701 & 1 \\
\hline
CWE-119 & 5 & CWE-362 & 1 & CWE-703 & 1 \\
\hline
CWE-125 & 3 & CWE-400 & 6 & CWE-754 & 1 \\
\hline
CWE-131 & 1 & CWE-415 & 2 & CWE-755 & 2 \\
\hline
CWE-134 & 1 & CWE-416 & 9 & CWE-758 & 1 \\
\hline
CWE-190 & 3 & CWE-427 & 2 & CWE-770 & 1 \\
\hline
CWE-191 & 1 & CWE-444 & 2 & CWE-787 & 3 \\
\hline
CWE-200 & 1 & CWE-475 & 1 & CWE-824 & 1 \\
\hline
CWE-203 & 3 & CWE-617 & 1 & CWE-908 & 1 \\
\hline
CWE-248 & 1 & CWE-662 & 1 &    &  \\
\hline
\end{tabular}
\label{table:CWEs}
\end{center}
\end{table}

\subsection{Research Question}

\textbf{RQ1: How effective is HALURust at detecting vulnerabilities in real-world Rust scenarios, and how do different prompts and LLMs impact its performance?}

Table \ref{table:ResultHALURust} presents the experimental results of HALURust with vulnerability analysis reports generated by different LLMs with various prompts. Overall, HALURust proves effective in detecting vulnerabilities within real-world Rust scenarios. Notably, HALURust paired with CO-STAR and Llama3 demonstrates the best performance, achieving the highest accuracy of 77.4\% (74.5\% + 2.9\%) and an F1 score of 77.3\% (74.8\% + 2.5\%). This performance stands out, especially when compared to the findings of Chakraborty et al. \cite{chakraborty2021deep}, who observed a performance decline of over 50\% in real-world vulnerability prediction using state-of-the-art deep learning methods and Ullah et al. \cite{ullah2024llms}, who concluded that current LLMs are not effective for vulnerability detection. The result reveals two significant insights:

\begin{table*}[h]
\caption{The Results of HALURust with Different LLMs (\%)}
\begin{center}
\begin{tabular}{|p{0.08\textwidth}|p{0.08\textwidth}|p{0.05\textwidth}|p{0.05\textwidth}|p{0.05\textwidth}|p{0.05\textwidth}|p{0.05\textwidth}|p{0.05\textwidth}|p{0.05\textwidth}|p{0.05\textwidth}|p{0.05\textwidth}|p{0.05\textwidth}|}
\hline
LLMs & Prompt &\multicolumn{2}{c|}{Accuracy} & \multicolumn{2}{c|}{Precision} & \multicolumn{2}{c|}{Recall} & \multicolumn{2}{c|}{F1 Score}\\
\hline
\multirow{2}{*}{GPT-3.5} & \multirow{2}{*}{TO} & \multirow{2}{*}{50.4} & +2.1 & \multirow{2}{*}{52.1} & +1.8 & \multirow{2}{*}{50.8} & +1.1 & \multirow{2}{*}{51.0} & +1.4  \\
& & & -5.5 &  & -4.1 &  & -4.7 &  &  -4.3\\
\hline
\multirow{2}{*}{Mistral} & \multirow{2}{*}{TO} & \multirow{2}{*}{43.4} & +6.7& \multirow{2}{*}{45.7} & +7.6 & \multirow{2}{*}{42.2} & +5.9 & \multirow{2}{*}{44.5} & +6.6  \\
& & & -9.3 &  & -8.9 &  & -8.3 &  &  -8.5\\
\hline
\multirow{2}{*}{Codellama} & \multirow{2}{*}{TO} & \multirow{2}{*}{50.3} & +3.7 & \multirow{2}{*}{51.2} & +2.9 & \multirow{2}{*}{39.7} & +4.3 & \multirow{2}{*}{45.4} & +3.3  \\
& & & -7.7 &  & -7.1 &  & -10.1 &  &  -8.4\\
\hline
\multirow{2}{*}{GPT-4o} & \multirow{2}{*}{TO} & \multirow{2}{*}{47.1} & +1.5 & \multirow{2}{*}{50.1} & +1.6 & \multirow{2}{*}{55.3} & +1.1 & \multirow{2}{*}{52.2} & +2.0  \\
& & & -4.9 &  & -2.8 &  & -1.7 &  &  -2.2\\
\hline
\multirow{2}{*}{Llama3} & \multirow{2}{*}{TO} & \multirow{2}{*}{51.5} & +3.6 & \multirow{2}{*}{54.4} & +2.7 & \multirow{2}{*}{52.1} & +3.3 & \multirow{2}{*}{53.8} & +3.1  \\
& & & -4.3 &  & -3.9 &  & -4.9 &  &  -4.4\\
\hline
\multirow{2}{*}{GPT-3.5} & \multirow{2}{*}{RO} & \multirow{2}{*}{56.4} & +1.4 & \multirow{2}{*}{58.7} & +3.5 & \multirow{2}{*}{66.7} & +2.1 & \multirow{2}{*}{62.5} & +2.7  \\
& & & -3.5 &  & -3.7 &  & -2.9 &  &  -3.4\\
\hline
\multirow{2}{*}{Mistral} & \multirow{2}{*}{RO} & \multirow{2}{*}{58.4} & +2.1& \multirow{2}{*}{58.9} & +1.9 & \multirow{2}{*}{74.7} & +1.1 & \multirow{2}{*}{66.3} & +1.5  \\
& & & -2.7 &  & -2.4 &  & -1.7 &  &  -1.9\\
\hline
\multirow{2}{*}{Codellama} & \multirow{2}{*}{RO} & \multirow{2}{*}{67.6} & +1.5 & \multirow{2}{*}{65} & +2.1 & \multirow{2}{*}{74.5} & +2.9 & \multirow{2}{*}{69.6} & +2.3  \\
& & & -2.1 &  & -3.1 &  & -2.3 &  &  -2.8\\
\hline
\multirow{2}{*}{GPT-4o} & \multirow{2}{*}{RO} & \multirow{2}{*}{54.1} & +3.7 & \multirow{2}{*}{56.7} & +2.9 & \multirow{2}{*}{61.8} & +1.5 & \multirow{2}{*}{59.1} & +2.2  \\
& & & -4.4 &  & -5.1 &  & -3.9 &  &  -4.4\\
\hline
\multirow{2}{*}{Llama3} & \multirow{2}{*}{RO} & \multirow{2}{*}{67.7} & +1.7 & \multirow{2}{*}{69.1} & +2.9 & \multirow{2}{*}{69.7} & +3.2 & \multirow{2}{*}{69.5} & +3.0  \\
& & & -3.3 &  & -2.7 &  & -2.9 &  &  -2.8\\
\hline
\multirow{2}{*}{GPT-3.5} & \multirow{2}{*}{CO-STAR} & \multirow{2}{*}{65.8} & +5.3 & \multirow{2}{*}{64.3} & +5.5 & \multirow{2}{*}{75.1} & +5.9 & \multirow{2}{*}{69.1} & +5.6  \\
& & & -5.4 &  & -6.1 &  & -4.8 &  &  -5.3\\
\hline
\multirow{2}{*}{Mistral} & \multirow{2}{*}{CO-STAR} & \multirow{2}{*}{69.4} & +2.7& \multirow{2}{*}{71.1} & +2.6 & \multirow{2}{*}{70.2} & +2.9 & \multirow{2}{*}{70.5} & +2.8  \\
& & & -5.3 &  & -5.9 &  & -6.1 &  &  -5.9\\
\hline
\multirow{2}{*}{Codellama} & \multirow{2}{*}{CO-STAR} & \multirow{2}{*}{69.7} & +2.5 & \multirow{2}{*}{70.4} & +2.8 & \multirow{2}{*}{73.2} & +1.9 & \multirow{2}{*}{71.6} & +2.1  \\
& & & -5.7 &  & -6.1 &  & -6.3 &  &  -6.2\\
\hline
\multirow{2}{*}{GPT-4o} & \multirow{2}{*}{CO-STAR} & \multirow{2}{*}{70.1} & +2.3 & \multirow{2}{*}{70.5} & +2.6 & \multirow{2}{*}{73.9} & +1.5 & \multirow{2}{*}{71.7} & +2.0  \\
& & & -4.9 &  & -7.1 &  & -5.9 &  &  -6.4\\
\hline
\multirow{2}{*}{Llama3} & \multirow{2}{*}{CO-STAR} & \multirow{2}{*}{\textbf {74.5}} & +2.9 & \multirow{2}{*}{\textbf {73.1}} & +2.6 & \multirow{2}{*}{\textbf {75.9}} & +3.8 & \multirow{2}{*}{\textbf {74.8}} & +2.5  \\
& & & -6.3 &  & -6.7 &  & -5.9 &  &  -6.1\\
\hline
\end{tabular}
\label{table:ResultHALURust}
\end{center}
\end{table*}

Firstly, the choice of prompt used for generating vulnerability reports significantly influences the performance of HALURust. Initially, using a task-oriented prompt that simply assigns a vulnerability detection task to the LLMs led to poor results, with accuracy and F1 scores ranging from 40\% to 55\%, barely surpassing random guessing. Introducing a role-oriented prompt by specifying the role of a "Rust security expert" yielded substantial performance gains across all LLMs. Notably, Codellama's accuracy rose from 50.3\% to 67.6\%, and its F1 score improved from 45.4\% to 69.6\%. This indicates that providing context through a specialized role significantly enhances the quality of the generated vulnerability reports, directly impacting HALURust's detection capabilities. Further refinement using the CO-STAR prompt, which imposes additional constraints on the report generation process, resulted in another performance boost. On average, accuracy and F1 scores increased by around 7\% across the LLMs, with GPT-4o showing the most notable improvement, exceeding a 10\% gain in both metrics. These findings underscore the critical role of prompt design in enhancing the effectiveness of HALURust, highlighting that well-structured and contextually rich prompts are essential for generating high-quality vulnerability analysis reports that drive accurate vulnerability detection.

Secondly, the performance of HALURust varies significantly depending on the LLM used to generate the reports. UUsing the best-performing CO-STAR prompt as an example, for all five LLMs, the average accuracy ranges from 65.8\% to 74.5\%, and the F1 score fluctuates between 69.1\% and 74.8\%. This variation underscores how different LLMs, using the same underlying code data, can yield diverse outcomes due to the distinct nature of the information each model generates. Then, even when using reports generated by the same LLM, there are considerable differences between the maximum positive and negative deviations across all measurements. This variability is influenced by the limited size and diversity of the data set.

Specifically, HALURust with Llama3 demonstrates superior performance compared to the other combinations, with an average accuracy of 74.5\% and an F1 score of 74.8\%, effectively handling both vulnerable and non-vulnerable samples. The close precision of 73.1\% and recall of 75.9\% indicate a balanced rate of false positives and false negatives, which are both manageable and acceptable. This enhanced performance is attributed to Llama3's use of a more extensive and updated corpus during training, resulting in reports that contain richer information. HALURust with Codellama and Mistral shows comparable performance, with both accuracy and F1 scores hovering around 70\%. The difference is that HALURust with Mistral achieves higher precision but lower recall compared to HALURust with Codellama. GPT-3.5 shows the most variability among these four LLMs. Its performance fluctuates significantly depending on the training and evaluation dataset used. The best performance for GPT-3.5 can reach an accuracy of 71.1\% (65.8\% + 5.3\%) and an F1 score of 74.7\% (69.1\% + 5.6\%), which are close to those of HALURust with Llama3. However, its worst performance is the lowest of all four combinations, and its average performance is slightly lower than Mistral and Codellama, and much lower than that of Llama3. As an advanced version of GPT-3.5, GPT-4o offers improved performance, though it still falls slightly short of Llama3. Based on these results, CO-STAR with Llama3 will be set as the default in HALURust for the upcoming experiments.

\textbf{RQ2: What characteristics make HALURust effective in detecting vulnerabilities in real-world Rust scenarios?}

\begin{figure*}[h]
\centering
\begin{tabular}{cc}
\includegraphics[scale=0.4]{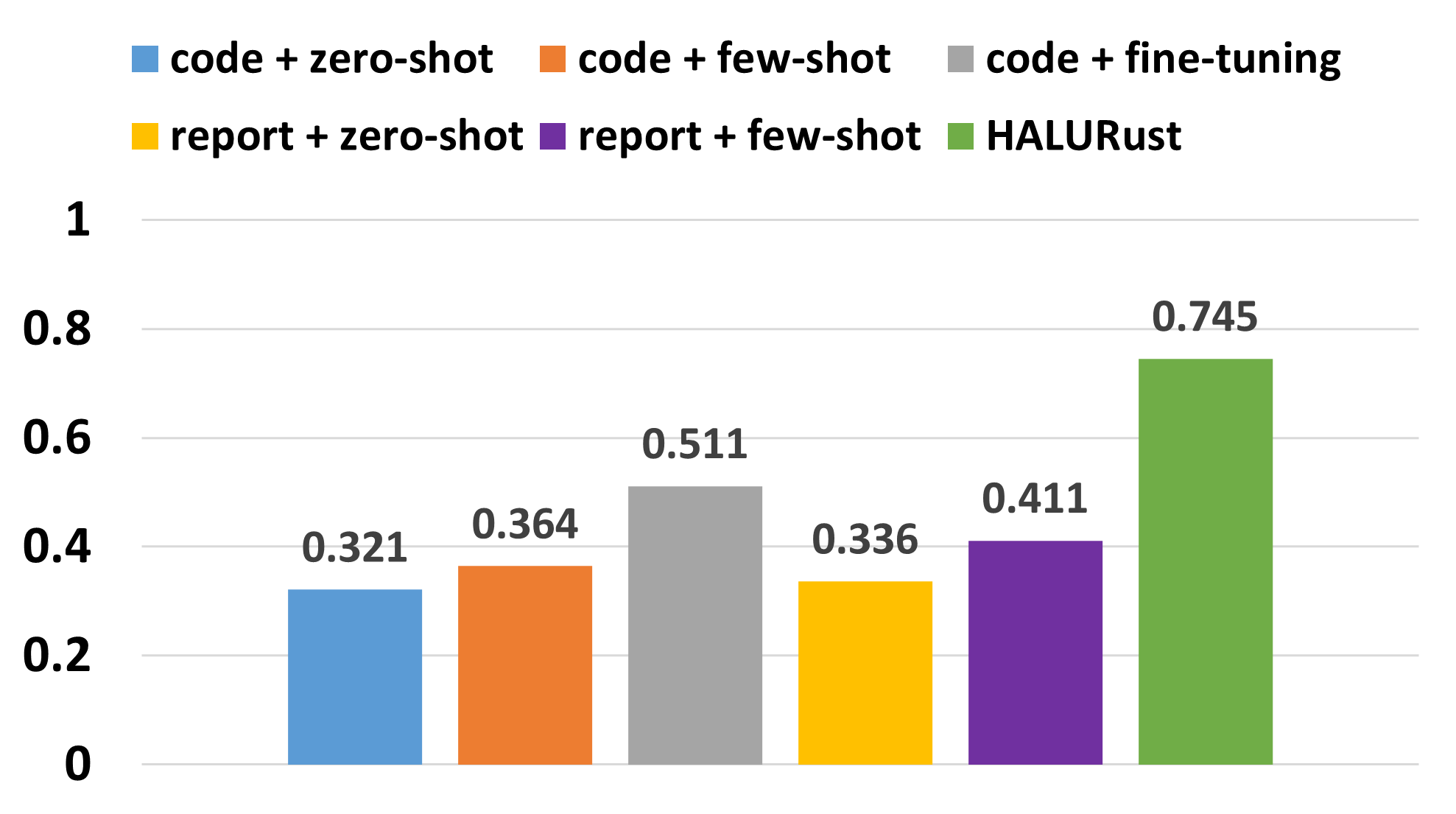}
&
\includegraphics[scale=0.4]{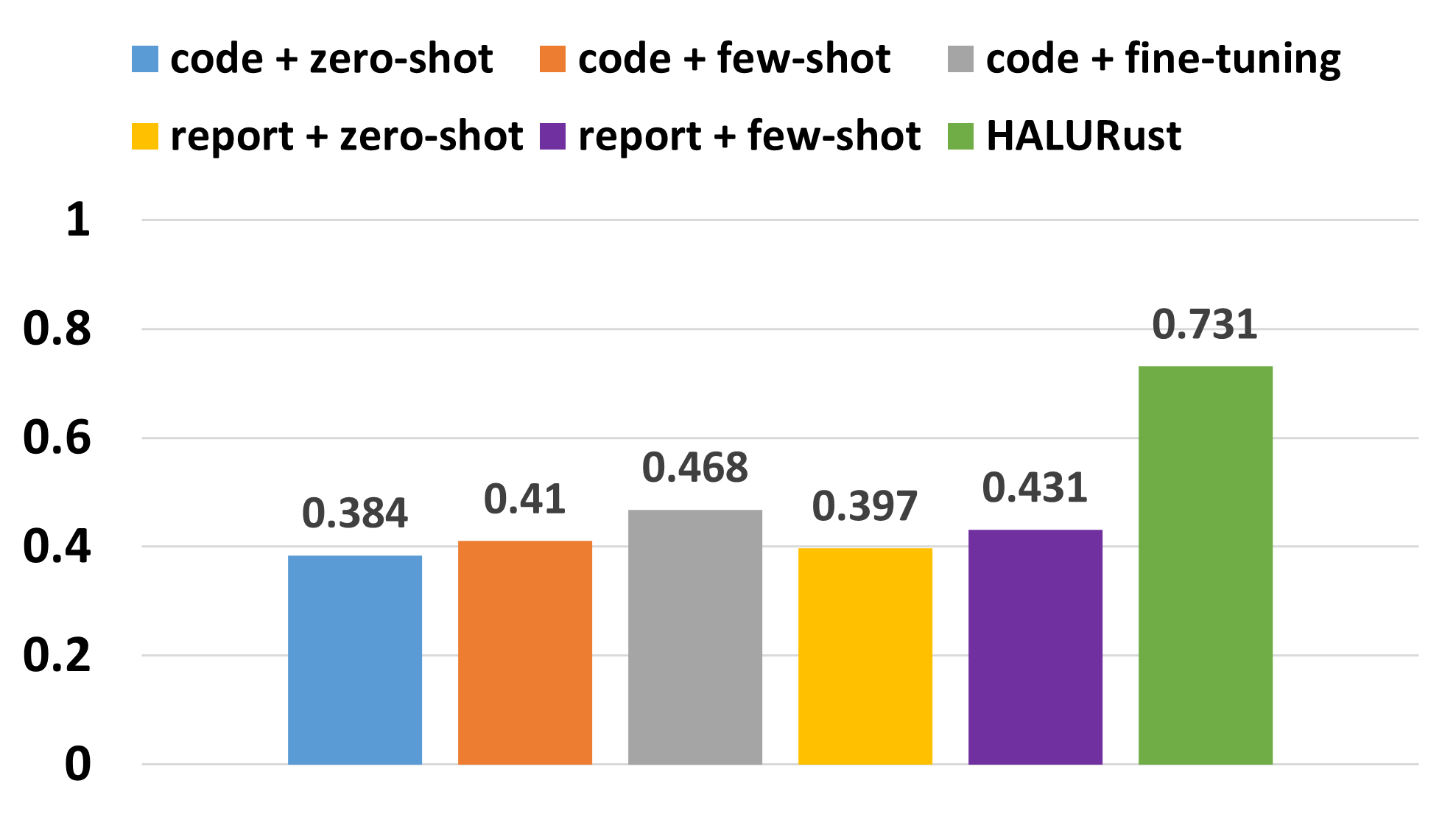} \\
(a) Accuracy  & (b) Precision\\
\includegraphics[scale=0.4]{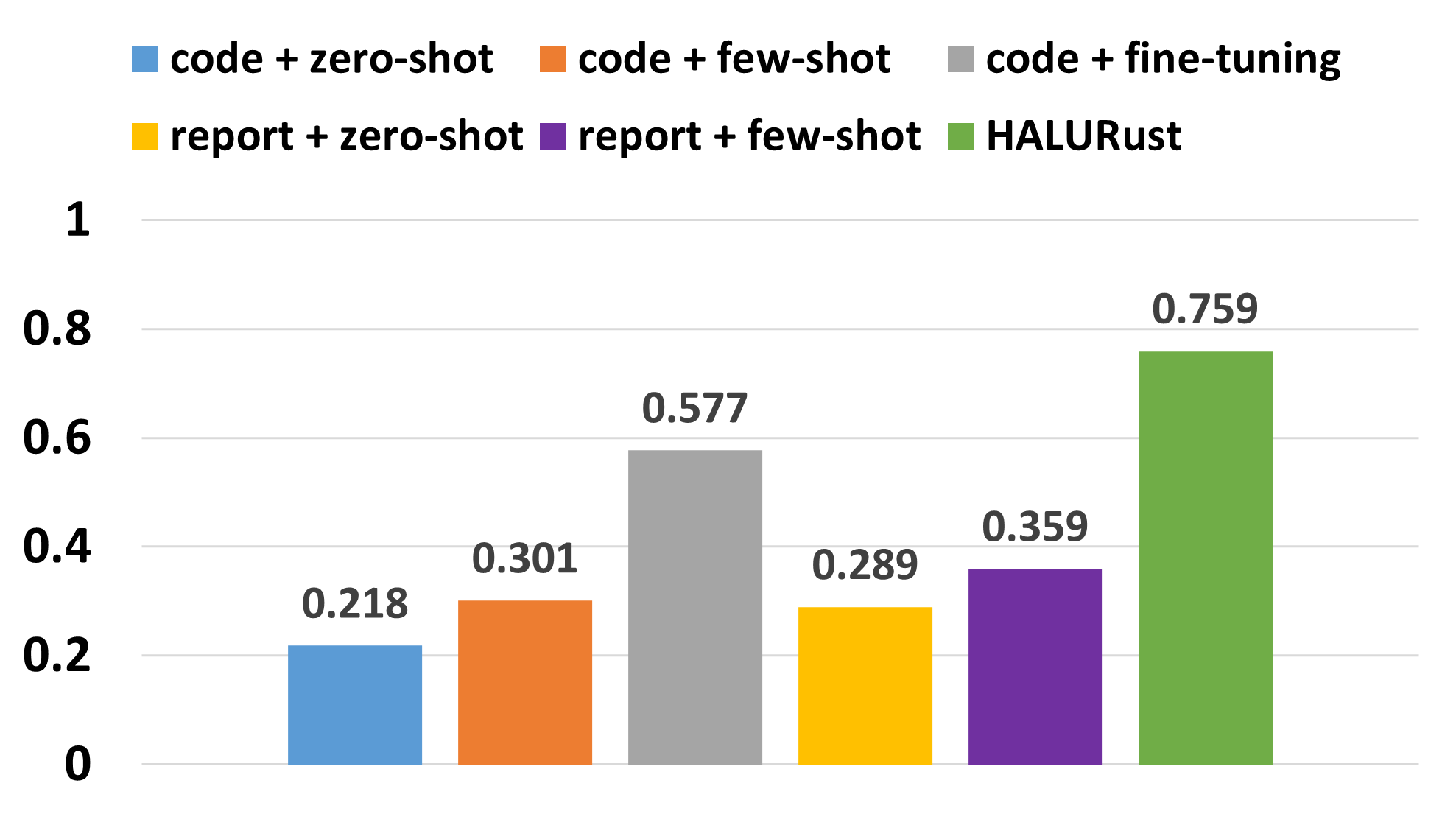} 
&
\includegraphics[scale=0.4]{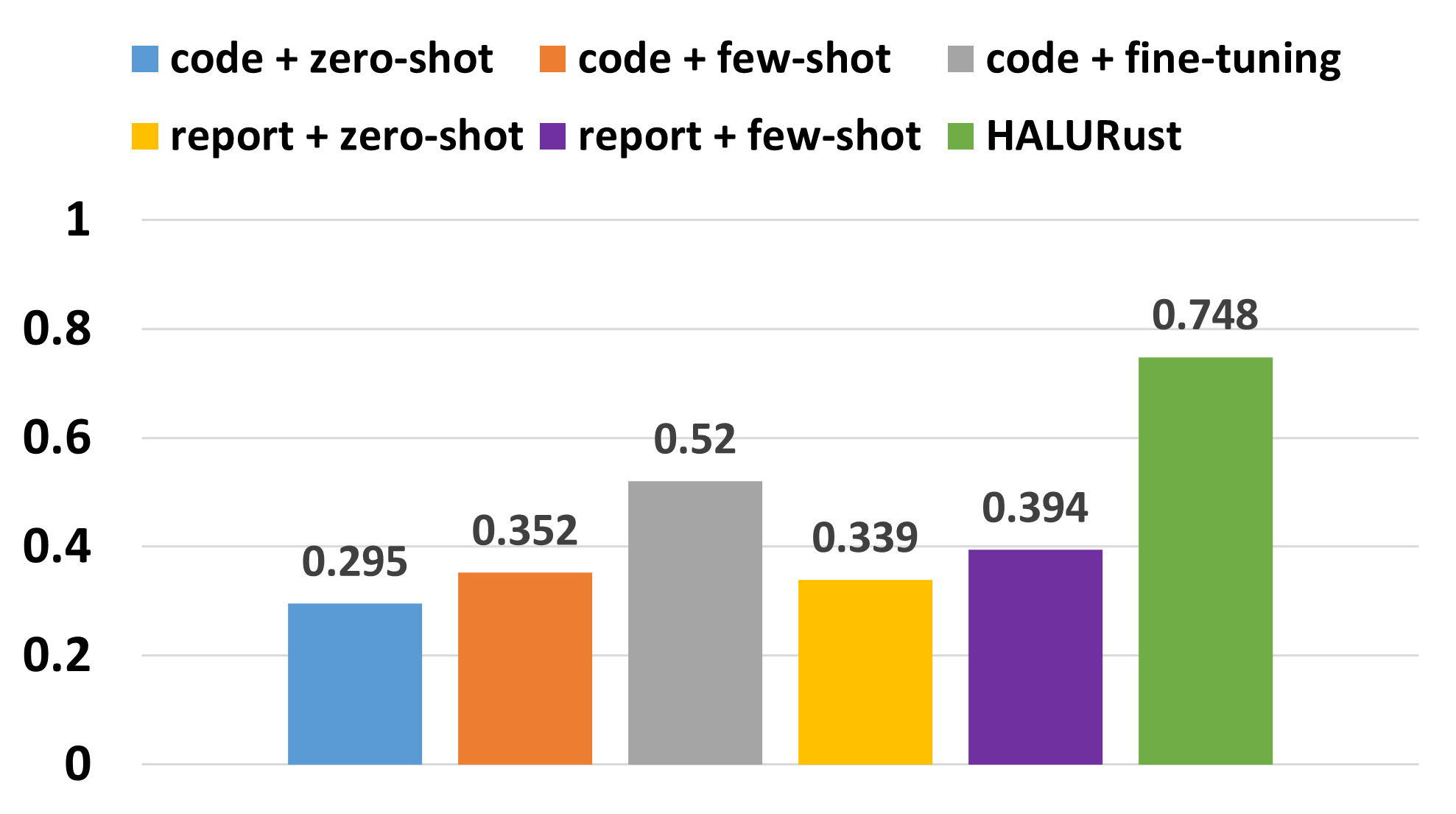}\\
(c) Recall  & (d) F1 Score\\
\end{tabular}
\caption{The Ablation Study on HALURust}
\label{fig:Ablation}
\end{figure*}

HALURust's effectiveness is attributed to two primary factors: (1) It utilizes vulnerability analysis reports as data inputs rather than raw source code, and (2) It applies 80\% of the samples in the dataset to fine-tune LLMs for specific vulnerability detection tasks. So, we address this research question by conducting ablation experiments on these two factors of HALURust.

Firstly, to evaluate the effectiveness of using report data as the format for inputs, we conducted an experiment where the source code directly serves as the input data, replacing vulnerability analysis reports. Secondly, to assess the impact of extensive fine-tuning, we modified the fine-tuning strategy by reducing the proportion of the dataset used: instead of fine-tuning with 80\% of the samples, we experimented with zero-shot and few-shot learning approaches. The experimental setups include: using code as input with zero-shot learning, code input with few-shot learning, code input with fine-tuning, report input with zero-shot learning, report input with few-shot learning, and HALURust.

The results of the ablation study are shown in Figure \ref{fig:Ablation}. Compared to other combinations, the HALURust framework consistently outperforms across all metrics. Firstly, both zero-shot and few-shot learning approaches, whether using direct source code inputs or vulnerability analysis reports, perform poorly. None of the models in these categories achieved an accuracy above 40\%, and the F1 scores were only slightly better, hovering around 40\%. These results are significantly lower compared to those obtained from models with extensive fine-tuning, underscoring the fact that LLMs generally require substantial data input to perform effectively.

Moreover, even with extensive fine-tuning, models using direct source code as input only reached about 50\% in both accuracy and F1 scores. Although the recall was somewhat higher at close to 60\%, indicating that 60\% of positive samples were correctly predicted, the models still missed many true positives, which could be critical in a real-world application. In summary, using vulnerability analysis reports as input and applying an extensive fine-tuning strategy are key factors that contribute to the superior performance of HALURust, with all four metrics approximately around 75\%.

\textbf {RQ3: To what extent does HALURust outperform other existing approaches in detecting vulnerabilities in real-world Rust scenarios?}

Detecting vulnerabilities in real-world scenarios is a significant challenge across all programming languages. Many state-of-the-art methods achieve as high as 90\% accuracy in controlled test environments, but their performance often drops by more than 50\% when applied in real-world settings, sometimes rendering the models no better than random guesses or even worse. In this experiment, we compare HALURust against 7 existing vulnerability detection methods: Devign \cite{zhou2019devign}, VulDeePecker \cite{Li2018}, MVD \cite{cao2022mvd}, D2A\cite{zheng2021d2a}, Linevul \cite{fu2022linevul} and tools developed by Le et al. \cite{le2024software} utilizing CodeBERT \cite{feng2020codebert} and ChatGPT \cite{openai2024chatgpt}. The first four approaches are language-agnostic, deep learning-based methods, primarily applied to the C language. Linevul uses a transformer-based architecture to capture both the syntactic and semantic information of code, enabling vulnerability detection at both the function and line levels. Le et al. focus on fine-tuning CodeBERT and ChatGPT with techniques like Random Over-Sampling (ROS) and Random Under-Sampling (RUS) to improve the models' performance in detecting vulnerabilities in low-resource languages.

\begin{figure}[h]
  \centering
  \includegraphics[scale = 0.33]{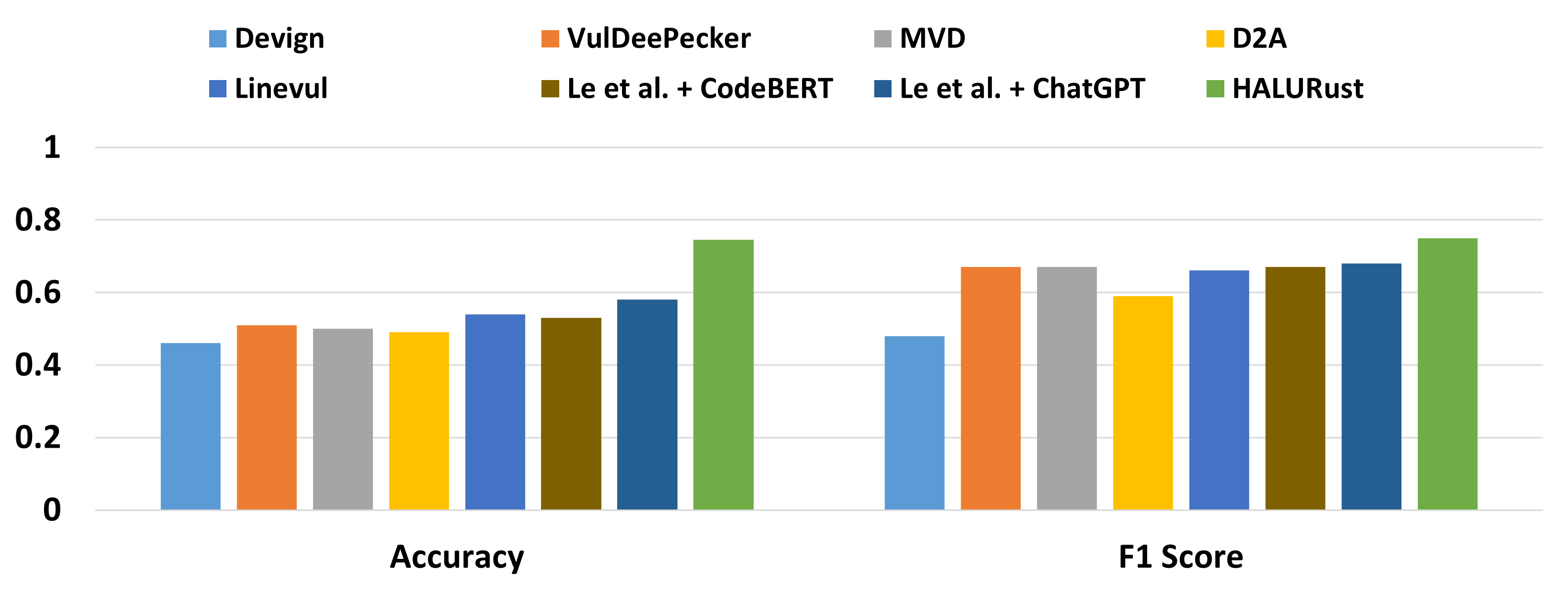}
  \caption{Comparison with Existing Methods}
  \label{fig: compare}
\end{figure}

Figure \ref{fig: compare} illustrates the accuracy and F1 score of each method, with a particular focus on their effectiveness in handling both positive and negative samples. The results highlight that HALURust significantly outperforms the other 7 methods in detecting vulnerabilities within real-world Rust scenarios. Specifically, HALURust achieves both accuracy and F1 scores close to 75\%, marking an approximate 10\% improvement over state-of-the-art methods. These metrics highlight HALURust's improved ability to accurately identify more true positives and true negatives while reducing errors. In contrast, the performance of the other seven methods is relatively similar, with accuracy ranging from 45\% to 55\% and F1 scores between 60\% and 70\%.

Generally, all 8 methods employ a similar training approach, utilizing extensive fine-tuning of transformer-based models. However, they differ in their sampling techniques and data input formats. HALURust’s superior performance can be attributed to its innovative use of LLMs' hallucinated outputs to discern vulnerabilities. This approach not only enhances the model’s accuracy in predicting vulnerabilities but also achieves a better balance between positive and negative samples, effectively reducing both false positives and false negatives compared to other models.

\textbf {RQ4: Can HALURust effectively handle unseen CWEs?}

\begin{table}[htbp]
\caption{Prediction Result on Unseen CWEs}
\begin{center}
\begin{tabular}{|c|c|c|c|c|c|}
\hline
\textbf{CWEs}& \textbf{Pred.} & \textbf{CWEs} & \textbf{Pred.} &\textbf{CWEs}& \textbf{Pred.}   \\
\hline
CWE-079 & \text{x} & CWE-287 & $\checkmark$ & CWE-668 & $\checkmark$ \\
\hline
CWE-088 & $\checkmark$ & CWE-288 & $\checkmark$  & CWE-674 & \text{x} \\
\hline
CWE-113 & \text{x} & CWE-346 & $\checkmark$ & CWE-701 & $\checkmark$ \\
\hline
CWE-131 & $\checkmark$ & CWE-347 & \text{x} & CWE-703 & $\checkmark$ \\
\hline
CWE-134 & $\checkmark$ & CWE-362 & $\checkmark$ & CWE-754 & \text{x} \\
\hline
CWE-191 & \text{x} & CWE-475 & \text{x} & CWE-758 & $\checkmark$ \\
\hline
CWE-200 & $\checkmark$ & CWE-617 & $\checkmark$ & CWE-770 & $\checkmark$ \\
\hline
CWE-248 & \text{x} & CWE-662 & \text{x} & CWE-824 & $\checkmark$ \\
\hline
CWE-276 & $\checkmark$ & CWE-665 & $\checkmark$ & CWE-908 & \text{x} \\
\hline
\multicolumn{6}{l}
{Pred. indicates whether the CWE is predictable or not.}
\end{tabular}
\label{table:CWEunseen}
\end{center}
\end{table}

The capability to identify unseen CWEs is crucial in real-world scenarios as it reflects the system's ability to address emerging and unknown types of vulnerabilities. To assess whether HALURust can effectively handle unseen CWEs, we divided the dataset into two groups based on the availability of positive samples for each CWE. The first group includes CWEs with more than one positive sample, which are used to fine-tune the model. The second group, considered unseen CWEs, comprises all CWEs with only one positive sample and is used to test HALURUSt's ability to identify vulnerabilities associated with these unseen CWEs. The experiments were conducted using binary classification, and the results are presented in Table \ref{table:CWEunseen}. In total, there are 27 CWEs categorized as unseen, out of which 17 were successfully identified while 10 were misclassified. The accuracy for unseen CWEs stands at about 63\%, which is slightly lower than HALURust's overall performance on all CWEs but still considered acceptable.

Specifically, HALURust, employing a transformer-based architecture, demonstrates a certain level of inferential ability when handling unseen CWEs, particularly for those with features that are closely related to what it has previously learned. However, for CWEs that possess non-related or even contrasting features compared to its training data, HALURust faces significant challenges in accurately identifying them. For instance, while HALURust may accurately identify CWE-190, which involves integers exceeding the maximum limit, it might misclassify CWE-191, where integers fall below the minimum limit, due to the contrasting nature of these errors. Additionally, HALURust fails to detect vulnerabilities that are significantly different from those in its training set, such as web-related (CWE-079) and cryptographic-related (CWE-347) vulnerabilities.

\textbf {RQ5: How effective is HALURust in adapting to other programming languages for vulnerability detection?}

To assess HALURust's effectiveness on different programming languages, we selected the 13 most commonly observed CWEs for both C and Java from the dataset introduced in \cite{luo2022cag}. For each CWE, 500 positive and 500 negative samples were chosen, resulting in a dataset of 6,500 positive and 6,500 negative samples for each language.

\begin{table}[htbp]
\caption{The Results of HALURust on C (\%)}
\begin{center}
\label{tab:CResult}
\begin{tabular}{|c|c|c|c|c|}
\hline
LLMs & Accuracy & Precision & Recall & F1 Score \\
\hline
GPT3.5 & 93.1 & 90.4 &  95.2 & 92.7\\ 
\hline
Mistral & 94.2 & 95.1 & 92.4 & 93.9\\ 
\hline
Codellama & 92.5 & 92.3 & 92.5 & 92.4 \\ 
\hline
GPT4 & 94.3 & 94.7 & 95.1 & 94.9\\ 
\hline
Llama3 & 94.7 & 95.1 & 9.48 & 95.0\\ 
\hline
\end{tabular}
\end{center}
\end{table}

\begin{table}[htbp]
\caption{The Results of HALURust on Java (\%)}
\begin{center}
\label{tab:JavaResult}
\begin{tabular}{|c|c|c|c|c|}
\hline
LLMs & Accuracy & Precision & Recall & F1 Score \\
\hline
GPT3.5 & 95.5 & 96.1 &  94.2 & 95.3 \\ 
\hline
Mistral & 96.6 & 97.2 & 95.5 & 96.3\\ 
\hline
Codellama & 94.7 & 95.7 & 95.3 & 95.5 \\ 
\hline
GPT4 & 96.1 & 95.8 &  97.1 & 96.4\\ 
\hline
Llama3 & 96.4 & 95.7 & 96.8 & 96.2 \\ 
\hline
\end{tabular}
\end{center}
\end{table}

Tabel \ref{tab:CResult} and \ref{tab:JavaResult} show the result of HALURust on C and Java, demonstrating strong adaptability across programming languages for vulnerability detection. For C, HALURust performs well, with accuracy scores ranging from 92.5\% to 94.7\% across different LLMs, and Llama3 achieves the highest F1 score of 95.0\%. In Java, performance further improves, with accuracy scores ranging from 94.7\% to 96.6\%. Mistral and Llama3 perform exceptionally well in Java, both achieving the highest F1 scores, indicating that HALURust adapts effectively to Java, showing a slight performance advantage over C.

\section{Threats to Validity}

\textbf{Potential Data Memorization.} The dataset includes real-world applications with 81 CVE records in Rust, which are openly available online. This raises the possibility that LLMs might have been exposed to some of this information during their training phase. However, the varying performance observed when using different prompts suggests that HALURust's effectiveness is not solely a result of data memorization by the LLMs.

\textbf{Threats to Data Quality.} HALURust relies on automatically generated vulnerability analysis reports to detect vulnerabilities, which could introduce potential errors. To address this, we manually verified all reports generated using the CO-STAR prompt and found no significant errors. While reports generated with the other two prompts were not manually checked, we ensured consistency across all samples and steps within HALURust. This uniformity minimizes the likelihood of performance impacts due to potential inaccuracies in the unchecked reports.

\textbf{Threats to Generalization.} Our dataset includes 44 CWEs, some of which have only one sample, potentially limiting the generalizability of the results. However, HALURust is not restricted to these 44 CWEs and can be extended to new and unseen CWEs in the future. RQ 4 further demonstrates HALURust's ability to handle unseen CWEs effectively.

\section{Conclusion}

We have developed the HALURust framework, which utilizes the hallucinatory outputs of LLMs to detect vulnerabilities in real-world Rust scenarios. With a peak accuracy of 77.4\% and an F1 score of 77.3\%, HALURust significantly surpasses existing state-of-the-art methods, achieving more than a 10\% improvement. Additionally, its robust performance on unseen types of CWEs highlights HALURust's capability to handle emerging and unknown vulnerabilities effectively.

While this paper constructs a dataset from real-world vulnerabilities across various applications, the dataset size remains limited. This limitation restricts the model's stability and presents challenges in predicting specific types of vulnerabilities accurately. Future work will focus on expanding the dataset to enhance the model's stability and adapt HALURust in predicting specific vulnerability types.

\clearpage

\bibliographystyle{IEEEtran} 
\bibliography{References}

\appendices

\section{CVE Information}

\begin{table*}[h]
  \caption{CVE Records (Part 1)}
  \label{tab:CVEInfo}
  \begin{tabular}{|l|l|l|l|l|}
    \hline
    CVE-ID & CWE Type & Program & Version & Reference \\
    \hline
    CVE-2015-20001 & CWE-119 & standard library in rust & before 1.2.0 & \href{https://github.com/rust-lang/rust/pull/25856/commits/5249cbb7fa31ea2e6e8d77b49bfda386215b1ce7}{5249cbb7fa31ea2e6e8d77b49bfda386215b1ce7} \\
    \hline
    CVE-2017-18016 & CWE-346 & Parity Browser & before 1.6.10 & \href{https://github.com/openethereum/parity-ethereum/commit/53609f703e2f1af76441344ac3b72811c726a215}{53609f703e2f1af76441344ac3b72811c726a215} \\
    \hline
    CVE-2018-1000657 & CWE-119 & standard library in rust & before 1.22.0 & \href{https://github.com/rust-lang/rust/commit/f71b37bc28326e272a37b938e835d4f99113eec2}{f71b37bc28326e272a37b938e835d4f99113eec2} \\
    \hline
    CVE-2018-1000810 & CWE-190 & standard library in rust & 1.26.0-1.29.0 & \href{https://github.com/rust-lang/rust/pull/54399/commits/8ac88d375e00c91a3db5d78852048322f88be3c1}{8ac88d375e00c91a3db5d78852048322f88be3c1} \\
    \hline
    CVE-2018-21000 & CWE-119 & safe-transmute crate & before 0.10.1 & \href{https://github.com/nabijaczleweli/safe-transmute-rs/pull/36/commits/a134e06d740f9d7c287f74c0af2cd06206774364}{a134e06d740f9d7c287f74c0af2cd06206774364} \\
    \hline
    CVE-2018-25008 & CWE-662 & standard library in rust & before 1.29.0 & \href{https://github.com/rust-lang/rust/pull/52031/commits/f96c2468695911222ba7557ce04af0dd8fbb6df2}{f96c2468695911222ba7557ce04af0dd8fbb6df2} \\
    \hline
    CVE-2019-1010299 & CWE-200 & standard library in rust & 1.18.0 and later & \href{https://github.com/rust-lang/rust/pull/53571/commits/b85e4cc8fadaabd41da5b9645c08c68b8f89908d}{b85e4cc8fadaabd41da5b9645c08c68b8f89908d} \\
    \hline
    CVE-2019-15541 & CWE-088 & rustls crate & before 0.16.0 & \href{https://github.com/rustls/rustls/commit/a93ee1abd2ab19ebe4bf9d684d56637ee54a6074}{a93ee1abd2ab19ebe4bf9d684d56637ee54a6074} \\
    \hline
    CVE-2019-15550 & CWE-125 & simd-json crate & before 0.1.1 & \href{https://github.com/simd-lite/simd-json/pull/27/commits/c838f93619498b741dfae1f86943c003ceb632ee}{c838f93619498b741dfae1f86943c003ceb632ee} \\
    \hline
    CVE-2019-16214 & CWE-701 & Libra Core & before 2019-09-0 & \href{https://github.com/diem/diem/commit/7efb0221989f17fdf7f8486730898ed947a1e19e}{7efb0221989f17fdf7f8486730898ed947a1e19e} \\
    \hline
    CVE-2019-16880 & CWE-415 & linea crate & 0.9.4 & \href{https://github.com/strake/linea.rs/pull/2/commits/dcacf864cbd2c3a3408d2c9010c6246dd732b969}{dcacf864cbd2c3a3408d2c9010c6246dd732b969} \\
    \hline
    CVE-2019-16882 & CWE-416 & string-interner crate & before 0.7.1 & \href{https://github.com/Robbepop/string-interner/pull/51/commits/33fcd548da205c1fa17409c51a16e221042a2104}{33fcd548da205c1fa17409c51a16e221042a2104} \\
    \hline
    CVE-2019-20399 & CWE-203 & Parity libsecp256k1-rs & before 0.3.1 & \href{https://github.com/paritytech/libsecp256k1/commit/11ba23a9766a5079918cd9f515bc100bc8164b50}{11ba23a9766a5079918cd9f515bc100bc8164b50} \\
    \hline
    CVE-2020-25573 & CWE-824 & linked-hash-map crate & before 0.5.3 & \href{https://github.com/contain-rs/linked-hash-map/pull/100/commits/df65b33f8a9dbd06b95b0a6af7521f0d47233545}{df65b33f8a9dbd06b95b0a6af7521f0d47233545} \\
    \hline
    CVE-2020-26297 & CWE-079 & mdBook & before 0.4.5 & \href{https://github.com/rust-lang/mdBook/commit/32abeef088e98327ca0dfccdad92e84afa9d2e9b}{32abeef088e98327ca0dfccdad92e84afa9d2e9b} \\
    \hline
    CVE-2020-35861 & CWE-125 & bumpalo crate & before 3.2.1 & \href{https://github.com/fitzgen/bumpalo/pull/70/commits/d08bc37de3dd471ea564a954a2de1d0caf045f65}{d08bc37de3dd471ea564a954a2de1d0caf045f65} \\
    \hline
    CVE-2020-35869 & CWE-134 & usqlite crate & before 0.23.0 & \href{https://github.com/rusqlite/rusqlite/commit/2327d3b774927fdf48903c0bdc1ca7ec93c7c8d0}{2327d3b774927fdf48903c0bdc1ca7ec93c7c8d0} \\
    \hline
    CVE-2020-35870 & CWE-416 & rusqlite crate & before 0.23.0 & \href{https://github.com/rusqlite/rusqlite/commit/2ef3628dac35aeba0a97d5fb3a57746b4e1d62b3}{2ef3628dac35aeba0a97d5fb3a57746b4e1d62b3} \\
    \hline
    CVE-2020-35904 & CWE-131 &  crossbeam-channel crate & before 0.4.4 & \href{https://github.com/crossbeam-rs/crossbeam/pull/533/commits/be327d581e8434a2ba41b74affc915f42d8abfcd}{be327d581e8434a2ba41b74affc915f42d8abfcd} \\
    \hline
    CVE-2020-35906 & CWE-416 & futures-task crate & before 0.3.6 & \href{https://github.com/rust-lang/futures-rs/pull/2206/commits/543687d6c85f39338a283d72614ea96a7fc81881}{543687d6c85f39338a283d72614ea96a7fc81881} \\
    \hline
    CVE-2020-35917 & CWE-416 & futures-task crate & before 0.3.6 & \href{https://github.com/PyO3/pyo3/pull/1297/commits/f86e6d392326f0e4dc25eeec8b26af36f2b0fbc8}{f86e6d392326f0e4dc25eeec8b26af36f2b0fbc8} \\
    \hline
    CVE-2020-35923 & CWE-416 &  ordered-float crate & before 1.1.1 & \href{https://github.com/reem/rust-ordered-float/pull/71/commits/7b5c8fe6b684239213e7e3e9d74c3dfc12599f16}{7b5c8fe6b684239213e7e3e9d74c3dfc12599f16} \\
    \hline
    CVE-2020-36210 & CWE-908 & autorand crate & before 0.2.3 & \href{https://github.com/mersinvald/autorand-rs/pull/7/commits/565d508993936821950009ec4c7c1e33301db81e}{565d508993936821950009ec4c7c1e33301db81e} \\
    \hline
    CVE-2020-36317 & CWE-787 & standard library in rust & before 1.49.0 & \href{https://github.com/rust-lang/rust/pull/78499/commits/e83666f45e3d93439775daefda7800b2ab193d30}{e83666f45e3d93439775daefda7800b2ab193d30} \\
    \hline
    CVE-2020-36318 & CWE-416 & standard library in rust & before 1.49.0 & \href{https://github.com/rust-lang/rust/pull/79814/commits/4fb9f1d7846f64beeec749db5933a24c05456ff2}{4fb9f1d7846f64beeec749db5933a24c05456ff2} \\
    \hline
    CVE-2021-21235 & CWE-400 & kamadak-exif & 0.5.2 & \href{https://github.com/kamadak/exif-rs/commit/f21df24616ea611c5d5d0e0e2f8042eb74d5ff48}{f21df24616ea611c5d5d0e0e2f8042eb74d5ff48} \\
    \hline
    CVE-2021-21299 & CWE-444 & hyper crate & 0.12.0-0.14.3 & \href{https://github.com/hyperium/hyper/commit/8f93123efef5c1361086688fe4f34c83c89cec02}{8f93123efef5c1361086688fe4f34c83c89cec02} \\
    \hline
CVE-2021-24117 & CWE-203 & Apache Teaclave Rust SGX SDK  & 1.1.3 & \href{https://github.com/dingelish/rust-base64/commit/a554b7ae880553db6dde8a387101a093911d5b2a}{a554b7ae880553db6dde8a387101a093911d5b2a} \\
\hline
CVE-2021-25900 & CWE-787 & smallvec crate & before 0.6.14 & \href{https://github.com/servo/rust-smallvec/pull/254/commits/9998ba0694a6b51aa6604748b00b6a98f0a0039e}{9998ba0694a6b51aa6604748b00b6a98f0a0039e} \\
\hline
CVE-2021-25902 & CWE-416 & glsl-layout crate & before 0.4.0 & \href{https://github.com/rustgd/glsl-layout/pull/10/commits/76bc9b5fbae73262307c41e72dbcfa0796073f30}{76bc9b5fbae73262307c41e72dbcfa0796073f30} \\
\hline
CVE-2021-28875 & CWE-252 & standard library in rust & before 1.50.0 & \href{https://github.com/rust-lang/rust/pull/80895/commits/ebe402dc9e708a8ed5e5860a7b30ea7826ab52a1}{ebe402dc9e708a8ed5e5860a7b30ea7826ab52a1} \\
\hline
CVE-2021-28876 & CWE-755 & standard library in rust & before 1.52.0 & \href{https://github.com/rust-lang/rust/pull/81741/commits/86a4b27475aab52b998c15f5758540697cc9cff0}{86a4b27475aab52b998c15f5758540697cc9cff0} \\
\hline
CVE-2021-28877 & CWE-119 & standard library in rust & before 1.51.0 & \href{https://github.com/rust-lang/rust/pull/80670/commits/af2983a9122138cb9055b79fda54e72f71599a6f}{af2983a9122138cb9055b79fda54e72f71599a6f} \\
\hline
CVE-2021-28878 & CWE-119 & standard library in rust & before 1.52.0 & \href{https://github.com/rust-lang/rust/pull/82292/commits/2371914a05f8f2763dffe6e2511d0870bcd6b461}{2371914a05f8f2763dffe6e2511d0870bcd6b461} \\
\hline
CVE-2021-28879 & CWE-190 & standard library in rust & before 1.52.0 & \href{https://github.com/rust-lang/rust/pull/82289/commits/66a260617a88ed1ad55a46f03c5a90d5ad3004d3}{66a260617a88ed1ad55a46f03c5a90d5ad3004d3} \\
\hline
CVE-2021-29511 & CWE-770 & evm & fix in >=0.21.1 & \href{https://github.com/rust-ethereum/evm/commit/19ade858c430ab13eb562764a870ac9f8506f8dd}{19ade858c430ab13eb562764a870ac9f8506f8dd} \\
\hline
CVE-2021-29922 & CWE-020 & standard library in rust & before 1.53.0 & \href{https://github.com/rust-lang/rust/pull/83652/commits/974192cd98b3efca8e5cd293f641f561e7487b30}{974192cd98b3efca8e5cd293f641f561e7487b30} \\
\hline
CVE-2021-31162 & CWE-415 & standard library in rust & before 1.52.0 & \href{https://github.com/rust-lang/rust/pull/83629/commits/421f5d282a51e130d3ca7c4524d8ad6753437da9}{421f5d282a51e130d3ca7c4524d8ad6753437da9} \\
\hline
CVE-2021-32715 & CWE-444 & hper crate & before 0.14.10 & \href{https://github.com/rust-lang/rust/pull/28826/commits/123a83326fb95366e94a3be1a74775df4db97739}{123a83326fb95366e94a3be1a74775df4db97739} \\
\hline
CVE-2021-32814 & CWE-022 & Skytable & prior to 0.5.1 & \href{https://github.com/skytable/skytable/commit/38b011273bb92b83c61053ae2fcd80aa9320315c}{38b011273bb92b83c61053ae2fcd80aa9320315c} \\
\hline
CVE-2021-37625 & CWE-252 & Skytable & prior to 0.6.4  & \href{https://github.com/skytable/skytable/commit/bb19d024ea1e5e0c9a3d75a9ee58ff03c70c7e5d}{bb19d024ea1e5e0c9a3d75a9ee58ff03c70c7e5d} \\
\hline
CVE-2021-36376 & CWE-427 & dandavison delta & before 0.8.3 & \href{https://github.com/dandavison/delta/commit/f01846bd443aaf92fdd5ac20f461beac3f6ee3fd}{f01846bd443aaf92fdd5ac20f461beac3f6ee3fd} \\
\hline
CVE-2021-36753 & CWE-427 & sharkdp BAT & before 0.18.2 & \href{https://github.com/sharkdp/bat/commit/bf2b2df9c9e218e35e5a38ce3d03cffb7c363956}{bf2b2df9c9e218e35e5a38ce3d03cffb7c363956} \\
\hline
CVE-2021-3917 & CWE-276 & coreos-installer & - & \href{https://github.com/coreos/coreos-installer/commit/2a36405339c87b16ed6c76e91ad5b76638fbdb0c}{2a36405339c87b16ed6c76e91ad5b76638fbdb0c} \\
\hline
CVE-2021-39193 & CWE-020 & Frontier & - & \href{https://github.com/polkadot-evm/frontier/commit/0b962f218f0cdd796dadfe26c3f09e68f7861b26}{0b962f218f0cdd796dadfe26c3f09e68f7861b26} \\
\hline
CVE-2021-41138 & CWE-020 & Frontier & - & \href{https://github.com/polkadot-evm/frontier/commit/146bb48849e5393004be5c88beefe76fdf009aba}{146bb48849e5393004be5c88beefe76fdf009aba} \\
\hline
CVE-2021-41153 & CWE-670 & evm crate & before 0.31.0 & \href{https://github.com/rust-ethereum/evm/pull/67/commits/a90cf79fcac7c8b56ee5301752938aa1d2e42609}{a90cf79fcac7c8b56ee5301752938aa1d2e42609} \\
\hline
CVE-2021-43790 & CWE-416 & Lucet & - & \href{https://github.com/bytecodealliance/lucet/commit/7c7757c772fb709c61b1442bcc1e1fbee97bf4a8}{7c7757c772fb709c61b1442bcc1e1fbee97bf4a8} \\
\hline
CVE-2021-44421 & CWE-203 & Occlum & before 0.26.0 & \href{https://github.com/occlum/occlum/commit/36918e42bf6732c4d3996bc99eb013eb6b90b249}{36918e42bf6732c4d3996bc99eb013eb6b90b249} \\

    \bottomrule
  \end{tabular}
\end{table*}

\begin{table*}
  \caption{CVE Records (Part 2)}
  \label{tab:CVEInfo2}
  \begin{tabular}{|l|l|l|l|l|}
    \hline
    CVE-ID & CWE Type & Program & Version & Reference \\
   \hline
    CVE-2022-21685 & CWE-191 & Frontier & - & \href{https://github.com/polkadot-evm/frontier/commit/8a93fdc6c9f4eb1d2f2a11b7ff1d12d70bf5a664}{8a93fdc6c9f4eb1d2f2a11b7ff1d12d70bf5a664} \\
    \hline
    CVE-2022-23639 & CWE-362 & crossbeam-utils & before 0.8.7 & \href{https://github.com/crossbeam-rs/crossbeam/pull/781/commits/f7c378b26e273d237575154800f6c2bd3bf20058}{f7c378b26e273d237575154800f6c2bd3bf20058} \\
    \hline
    CVE-2022-23066 & CWE-682 & Solana rBPF & 0.2.26,0.2.27 & \href{https://github.com/solana-labs/rbpf/commit/e61e045f8c244de978401d186dcfd50838817297}{e61e045f8c244de978401d186dcfd50838817297} \\
    \hline
    CVE-2022-24713 & CWE-400 & regex & <=1.5.4  & \href{https://github.com/rust-lang/regex/commit/ae70b41d4f46641dbc45c7a4f87954aea356283e}{ae70b41d4f46641dbc45c7a4f87954aea356283e} \\
    \hline
    CVE-2022-24791 & CWE-416 & Wasmtime & >0.34.2 & \href{https://github.com/bytecodealliance/wasmtime/commit/666c2554ea0e1728c35aa41178cf235920db888a}{666c2554ea0e1728c35aa41178cf235920db888a} \\
    \hline
    CVE-2022-27815 & CWE-059 & SWHKD & 1.1.5 & \href{https://github.com/waycrate/swhkd/commit/e661a4940df78fbb7b52c622ac4ae6a3a7f7d8aa}{e661a4940df78fbb7b52c622ac4ae6a3a7f7d8aa} \\
    \hline
    CVE-2022-27816 & CWE-059 & SWHKD & 1.1.5 & \href{https://github.com/waycrate/swhkd/commit/0b620a09605afb815c6d8d8953bbb7a10a8c0575}{0b620a09605afb815c6d8d8953bbb7a10a8c0575} \\
    \hline
    CVE-2022-27818 & CWE-668 & SWHKD & 1.1.5 & \href{https://github.com/waycrate/swhkd/commit/f70b99dd575fab79d8a942111a6980431f006818}{f70b99dd575fab79d8a942111a6980431f006818} \\
    \hline
    CVE-2022-27819 & CWE-400 & SWHKD & 1.1.5 & \href{https://github.com/waycrate/swhkd/commit/b4e6dc76f4845ab03104187a42ac6d1bbc1e0021}{b4e6dc76f4845ab03104187a42ac6d1bbc1e0021} \\
    \hline
    CVE-2022-31099 & CWE-674 & rulex & <0.4.3 & \href{https://github.com/pomsky-lang/pomsky/commit/60aa2dc03a22d69c8800fec81f99c96958a11363}{60aa2dc03a22d69c8800fec81f99c96958a11363} \\
    \hline
    CVE-2022-31100 & CWE-617 & rulex & <0.4.3 & \href{https://github.com/pomsky-lang/pomsky/commit/fac6d58b25c6f9f8c0a6cdc4dec75b37b219f1d6}{fac6d58b25c6f9f8c0a6cdc4dec75b37b219f1d6} \\
    \hline
    CVE-2022-31111 & CWE-670 & Frontier & - & \href{https://github.com/polkadot-evm/frontier/commit/e3e427fa2e5d1200a784679f8015d4774cedc934}{e3e427fa2e5d1200a784679f8015d4774cedc934} \\
    \hline
    CVE-2022-31169 & CWE-682 & Wasmtime & prior to 0.38.2 & \href{https://github.com/bytecodealliance/wasmtime/commit/2ba4bce5cc719e5a74e571a534424614e62ecc41}{2ba4bce5cc719e5a74e571a534424614e62ecc41} \\
    \hline
    CVE-2022-31173 & CWE-400 & Juniper & <0.15.10 & \href{https://github.com/graphql-rust/juniper/commit/2b609ee057be950e3454b69fadc431d120e407bb}{2b609ee057be950e3454b69fadc431d120e407bb} \\
    \hline
    CVE-2022-35922 & CWE-400 & Rust-WebSocket & prior to 0.26.5 & \href{https://github.com/websockets-rs/rust-websocket/commit/cbf6e9983e839d2ecad86de8cd1b3f20ed43390b}{cbf6e9983e839d2ecad86de8cd1b3f20ed43390b} \\
    \hline
    CVE-2022-36008 & CWE-190 & Frontier & - & \href{https://github.com/polkadot-evm/frontier/commit/fff8cc43b7756ce3979a38fc473f38e6e24ac451}{fff8cc43b7756ce3979a38fc473f38e6e24ac451} \\
    \hline
    CVE-2022-36113 & CWE-022 & Cargo & before 1.64 & \href{https://github.com/rust-lang/cargo/commit/97b80919e404b0768ea31ae329c3b4da54bed05a}{97b80919e404b0768ea31ae329c3b4da54bed05a} \\
    \hline
    CVE-2022-39252 & CWE-287 & standard library in rust & before 0.6 & \href{https://github.com/matrix-org/matrix-rust-sdk/commit/41449d2cc360e347f5d4e1c154ec1e3185f11acd}{41449d2cc360e347f5d4e1c154ec1e3185f11acd} \\
    \hline
    CVE-2023-22466 & CWE-665 & Tokio & 1.18.4,1.20.3,1.23.1 & \href{https://github.com/tokio-rs/tokio/pull/5336/commits/9ca156c0b80347bb7b1406c6050d602f22efc709}{9ca156c0b80347bb7b1406c6050d602f22efc709} \\
    \hline
    CVE-2023-28113 & CWE-347 & russh & >0.34.0;<0.37.1 & \href{https://github.com/warp-tech/russh/commit/d831a3716d3719dc76f091fcea9d94bd4ef97c6e}{d831a3716d3719dc76f091fcea9d94bd4ef97c6e} \\
    \hline
    CVE-2023-30624 & CWE-758 & Wasmtime & before 6.0.2 & \href{https://github.com/bytecodealliance/wasmtime/commit/0977952dcd9d482bff7c288868ccb52769b3a92e}{0977952dcd9d482bff7c288868ccb52769b3a92e} \\
    \hline
    CVE-2023-41051 & CWE-125 & vm-memory rust crate & > 0.1.0,< 0.12.2  & \href{https://github.com/rust-vmm/vm-memory/commit/aff1dd4a5259f7deba56692840f7a2d9ca34c9c8}{aff1dd4a5259f7deba56692840f7a2d9ca34c9c8} \\
    \hline
    CVE-2023-41317 & CWE-755 & Apollo Router & before v1.29.1 & \href{https://github.com/apollographql/router/commit/b295c103dd86c57c848397d32e8094edfa8502aa}{b295c103dd86c57c848397d32e8094edfa8502aa} \\
    \hline
    CVE-2023-45812 & CWE-754 & Apollo Router & - & \href{https://github.com/apollographql/router/pull/4014/commits/bfd17efb1bd3183c4a7b4e61280ee9769befd1e6}{bfd17efb1bd3183c4a7b4e61280ee9769befd1e6} \\
    \hline
    CVE-2023-46135 & CWE-248 & rs-stellar-strkey & <0.0.8 & \href{https://github.com/stellar/rs-stellar-strkey/pull/59/commits/f930025ff78ab0cfd35af8ab7d13793e12d5ed1b}{f930025ff78ab0cfd35af8ab7d13793e12d5ed1b} \\
    \hline
    CVE-2023-50711 & CWE-787 & vmm-sys-util & >0.5.0,<0.12.0 & \href{https://github.com/rust-vmm/vmm-sys-util/commit/30172fca2a8e0a38667d934ee56682247e13f167}{30172fca2a8e0a38667d934ee56682247e13f167} \\
    \hline
    CVE-2023-6245 & CWE-020 & Candid & - & \href{https://github.com/dfinity/candid/pull/478/commits/0d4e122f28715a81d17bcbb72a192e97224fe1ab}{0d4e122f28715a81d17bcbb72a192e97224fe1ab} \\
    \hline
    CVE-2024-21491 & CWE-288 & svix & before 1.17.0 & \href{https://github.com/svix/svix-webhooks/commit/958821bd3b956d1436af65f70a0964d4ffb7daf6}{958821bd3b956d1436af65f70a0964d4ffb7daf6} \\
    \hline
    CVE-2024-21629 & CWE-703 & Rust EVM & 0.41.1 & \href{https://github.com/rust-ethereum/evm/commit/d8991ec727ad0fb64fe9957a3cd307387a6701e4}{d8991ec727ad0fb64fe9957a3cd307387a6701e4} \\
    \hline
    CVE-2024-23644 & CWE-113 & Trillium & <0.3.12 & \href{https://github.com/trillium-rs/trillium/commit/16a42b3f8378a3fa4e61ece3e3e37e6a530df51d}{16a42b3f8378a3fa4e61ece3e3e37e6a530df51d} \\
    \hline
    CVE-2024-28854 & CWE-400 & tls-listener & prior to 0.10.0 & \href{https://github.com/tmccombs/tls-listener/commit/d5a7655d6ea9e53ab57c3013092c5576da964bc4}{d5a7655d6ea9e53ab57c3013092c5576da964bc4} \\
    \hline
    CVE-2024-20380 & CWE-475 & ClamAV & - & \href{https://github.com/Cisco-Talos/clamav/pull/1242/commits/af4270bfc7e7818eb03d06ca2d52c868cd373358}{af4270bfc7e7818eb03d06ca2d52c868cd373358} \\
    \bottomrule
  \end{tabular}
\end{table*}

\end{document}